\documentclass[proof]{pasj01}
\usepackage{multirow}
\usepackage{graphicx}
\Received{$\langle$reception date$\rangle$}
\Accepted{$\langle$acception date$\rangle$}
\Published{$\langle$publication date$\rangle$}
\begin{document}

\title{{ 
ALMA view of the Galactic super star cluster RCW\,38 at 270-AU resolution
}}
\author{Kazufumi Torii\altaffilmark{1}, Kazuki, Tokuda\altaffilmark{2,3}, Kengo Tachihara\altaffilmark{4}, Toshikazu Onishi\altaffilmark{2}, Yasuo Fukui\altaffilmark{4,5}}%
\altaffiltext{1}{Nobeyama Radio Observatory, National Astronomical Observatory of Japan (NAOJ), National Institutes of Natural Sciences (NINS), 462-2, Nobeyama, Minamimaki, Minamisaku, Nagano 384-1305, Japan}
\altaffiltext{2}{Department of Physical Science, Graduate School of Science, Osaka Prefecture University, 1-1 Gakuen-cho, Naka-ku, Sakai, Osaka 599-8531, Japan }
\altaffiltext{3}{Chile Observatory, National Astronomical Observatory of Japan, National Institutes of Natural Science, 2-21-1 Osawa, Mitaka, Tokyo 181-8588, Japan}
\altaffiltext{4}{Graduate School of Science, Nagoya University, Chikusa-ku, Nagoya, Aichi 464-8601, Japan}
\altaffiltext{5}{Institute for Advanced Research, Nagoya University, Furo-cho, Chikusa-ku, Nagoya 464-8601, Japan}

\email{tokuda@p.s.osakafu-u.ac.jp}

\KeyWords{ISM: clouds --- ISM: molecules --- radio lines: ISM --- stars: formation}

\maketitle

\begin{abstract}
We report millimeter/submillimeter continuum and molecular line observations of the Galactic super star cluster RCW\,38, obtained from the Atacama Large Millimeter/Submillimeter Array with a minimum angular resolution of $0''17\times0''15$ ($\simeq289\,{\rm AU}\times255\,{\rm AU}$). 
The C$^{18}$O image reveal many massive condensations embedded within filamentary structures extending along the northwest-southeast direction in the center of cluster. The condensations have sizes of 0.01--0.02\,pc, H$_2$ column densities of $10^{23}$--$10^{24}$\,cm$^{-2}$, and H$_2$ masses of 10--130\,$M_\odot$.
In addition, the 233-GHz continuum image reveals two dense, small millimeter-sources with radii of 460 and 200\,AU (Source\,A and Source\,B). 
Source\,A is embedded within the most massive C$^{18}$O condensation, whereas no counterpart is seen for Source\,B.
The masses of Source\,A and Source\,B are estimated as 13 and 3\,$M_\odot$ at the optically-thin limit, respectively. 
The C$^{18}$O emission shows a velocity gradient of 2\,km\,s$^{-1}$ at the central 2000\,AU of Source\,A, which could be interpreted as a Keplerian rotation with a central mass of a few $M_\odot$ or infall motion of gas. 
Further, the ALMA $^{12}$CO data reveal that Source\,A and Source\,B are associated with molecular outflows exhibiting maximum velocities of $\sim$30--70\,km\,s$^{-1}$.
The outflows have short dynamical timescales of $<$1000\,yr and high mass outflow rates of $\sim10^{-4}$--$10^{-3}$\,$M_\odot$\,yr$^{-1}$. These observational signatures suggest an early evolutionary phase of the massive star formation in Source\,A and Source\,B.
\end{abstract}

\section{Introduction}
The formation mechanism of massive stars is not yet thoroughly understood.
A number of theoretical models invoke a scaled-up version of the low-mass model, with a disk-like structure and jets or outflows (e.g., \cite{kur2007,kur2009,kur2010,kui2010,kui2011, kui2018, pet2010a,hos2009, hos2010}). Otherwise, massive young stellar objects (YSOs) could not accrete by overcoming the powerful radiation pressure of luminous protostars (e.g., \cite{wol1986, nak1989}).
Observations of jets and outflows associated with massive YSOs, which resemble those of low-mass YSOs, provide indirect evidence of high-mass accretion disks \citep{beu2002, mau2015a}.
However, observations providing a 100\,AU-scale resolution of the structures of massive YSOs are scarce. 

Interferometric observations with an angular resolution of a few hundred AU can be anticipated to reveal fragmentation, collapse, and outflowing gas of hidden dense materials that lead to the protostar formation, as demonstrated by recent Atacama Large Millimeter/submillimeter Array (ALMA) observations in nearby low- and high-mass star-forming regions (e.g., \cite{tok2014, tok2016, tok2019b, hir2017, oha2018, cas2019, mat2019}).
Several interferometric centimeter to submillimeter studies of luminous massive YSOs have resulted in the discovery of rotating structures on the order of of 100 to 1000\,AU \citep{san2013, guz2014, joh2015, ces2017, beu2017, mot2019}.
However, it is still crucial to study a greater number of massive YSOs resolved at the 100 to 1000\,AU scale.

In this paper, we report new observations of the Galactic super star cluster RCW\,38, obtained using the ALMA cycle-3 capability.
RCW\,38 is an outstanding super star cluster located at 1.7\,kpc, and harbors $\sim10^4$ stars, including $\sim$20 O-type stars (and candidates) within a small region of a few parsecs (\cite{wol2006, win2011, kuh2015}, see also a review by \cite{wol2008}). 
Most importantly, RCW\,38 is the youngest super star cluster reported in the Milky Way, with an age of only $\sim$0.1--0.5\,Myrs \citep{fuk2016, wol2008}. 
Despite strong dissipative and destructive effects caused by feedback from the formed massive stars, plenty of molecular gas still remains within the central $\sim$1\,pc of RCW\,38, offering a unique opportunity to observe the natal gas of the super star cluster during formation \citep{fuk2016}.

RCW\,38 has two sources that are bright in the near-infrared. 
The brightest feature at 2\,$\mu$m is known as IRS\,2, an O5.5 binary located at the center of the cluster \citep{der2009}, whose luminosity was measured as $7\times10^5$\,$L_\odot$ \citep{fur1975}.
Another bright feature, labeled IRS\,1, is located 0.1\,pc west of IRS\,2; this feature is a dust ridge extending by 0.1--0.2\,pc in the north-south direction.
The central 0.1\,pc of IRS\,1 and IRS\,2 is devoid of dust, with cavities evidenced by infrared and millimeter observations \citep{huc1974, vig2004, wol2006, wol2008}.

Based on molecular line observations acquired with single-dish millimeter and submillimeter telescopes, \citet{fuk2016} claimed that the formation of O stars in RCW\,38 was triggered by a collision between two molecular clouds.
The two identified clouds have radial velocities of $\sim$2 and $\sim$12\,km\,s$^{-1}$. 
The former ``the ring cloud'', which has a ring-like structure surrounding the cavities of IRS\,1 and IRS\,2, has a mass of $3\times10^4$\,$M_\odot$ and is the primary body of the natal cloud of RCW\,38, whereas the latter ``the finger cloud'', which is elongated across the cluster, is less dense, with a mass of $2\times10^3$\,$M_\odot$.
The two clouds are connected in velocity space by an intermediate velocity feature (called ``a bridge feature'').
The bridge feature is located at 0.3\,pc north of IRS\,1 and IRS\,2, indicating that the collisional interaction still continues, as suggested by the cloud-cloud collision model \citep{tor2015,tor2017,fuk2018}.
\citet{izu2019_pdf} observed these molecular clouds in the [C{\sc i}] emission line, indicating that the abundance ratio of CO to C{\sc i} is constant for visual extinction $A_{\rm v}$ of a few mag to 100 mag. 
This result may be interpreted with the clumpy photodissociation region (PDR) model (e.g., \cite{spa1997,tac2018}), in which clumpy structures with sizes of less than a few 0.1\,pc are predicted in the molecular clouds.

The ALMA observations in RCW\,38 (P.I.: Fukui,~Y., \#2015.1.01134.S) were conducted for the central 1.5\,pc\,$\times$\,0.7\,pc of IRS\,1 and IRS\,2 in the ALMA band-6 and band-7, using the main 12-m array and the Atacama Compact Array (ACA).
The minimum angular resolution of the dataset is $0''.17\times0''.15$, which corresponds to 289$\times$255\,AU at 1.7\,kpc.
%This high angular resolution allowed us to detect two dense sources of $\sim$500--1000\,AU in the central region of RCW\,38 with a small size on the order of 0.1\,pc.
The sources are embedded within dense dust condensations and are associated with molecular outflows having high mass outflow rates, suggesting massive star formation.
The remainder of this paper is organized as follows.
Section\,2 describes the ALMA dataset used in this study. 
Section\,3 presents the main results obtained by analyzing the ALMA dataset. 
The results are discussed in Section\,4, and finally, a summary is presented in Section\,5.

\section{Observations}
The parameters of the ALMA observations are summarized in Table\,\ref{tab:1}, whereas the datasets used in this study are listed in Table\,\ref{tab:2}.
The ALMA observations were conducted using the cycle-3 band-6 (1.3\,mm)  and band-7 (0.87\,mm) capabilities. 
The band-6 observations were performed using the 12-m array and the ACA (the 7-m array and Total Power (TP) array). 
The target molecular lines were $^{12}$CO $J$=2--1, $^{13}$CO $J$=2--1, C$^{18}$O $J$=2--1, SiO $J$=5--4, and H30$\alpha$, with spectral resolutions set to 141\,kHz ($\sim$0.18\,km\,s$^{-1}$) or 244\,kHz ($\sim$0.31\,km\,s$^{-1}$). 
A baseband with a bandwidth of 1875\,MHz centered on 233\,GHz was used to obtain the continuum emission. The pointing numbers of the 12-m array with the C40-6 configuration and the 7-m array were 33 and 13, respectively. The projected baselines of the 12-m and 7-m array observations were 15--3247\,m and 8--44\,m, respectively. The bandpass, flux, and phase calibrators were J1107-4449, J1107-4449, and J0851-5228, respectively, in the 12-m array observations. With the 7-m array, they ware J1058+0133, Callisto, and J0904-5735, respectively.

The band-7 observations, on the other hand, were carried out with the ACA alone. 
Three spectral windows were used to observe molecular lines of $^{13}$CO $J$=3--2, C$^{18}$O $J$=3--2, and CS $J$=7--6 at a spectral resolution of 244\, kHz ($\sim$0.22\,km\,s$^{-1}$), while the continuum emission was observed using a baseband with a bandwidth of 1875 MHz centered on 341\,GHz. The pointing number of the mosaicing observation was 27. The projected baseline length of the 7-m array ranged form 8 to 44\,m. We observed J1256-0547 or J0522-3627 as the bandpass calibrators, and J1256-0547 or J0522-3627 as the flux calibrators.
The phase calibrator was J0845-5458 throughout the observations.

In this study, only the datasets of CO lines and continuum emission were used to investigate the several 100 to 1000 AU-scale structures of RCW\,38.
Herein, individual datasets are denoted by the names listed in the ``Name'' column in Table\,\ref{tab:2}.
The obtained data were reduced using the Common Astronomy Software Application (CASA) package \citep{mcm2007}, whose version is 5.5.0. We used the  \texttt{tclean} task using the Briggs weighting with a robust parameter of 0.5.
In order to acquire better sensitivities, a uv-tapering was applied to the $^{12}$CO and C$^{18}$O$_{\rm all}$ data, which resulted in larger synthesized beams of $\sim$0$''.$5 and $\sim$2$''$, respectively. For the C$^{18}$O $J$=2--1 emission, two data cubes with different beam sizes (C$^{18}$O$_{\rm all}$ and C$^{18}$O) were generated to illustrate the large-scale gas structures and the small-scale gas kinematics around compact mm-sources in RCW\,38 (Figures\,\ref{fig1} and \ref{c18o}).
Note that The 12-m array and 7-m data were combined together in visibility space to make the combined image, and then we used the \texttt{feathering} task in CASA to merge the 12-m + 7-m array and the TP data for the C$^{18}$O$_{\rm all}$ data.%, whereas only the 12-m array data was used for the C$^{18}$O data.

\section{Results}
\subsection{Large-scale C$^{18}$O distribution}
\begin{figure}
 \begin{center}
  \includegraphics[width=15cm]{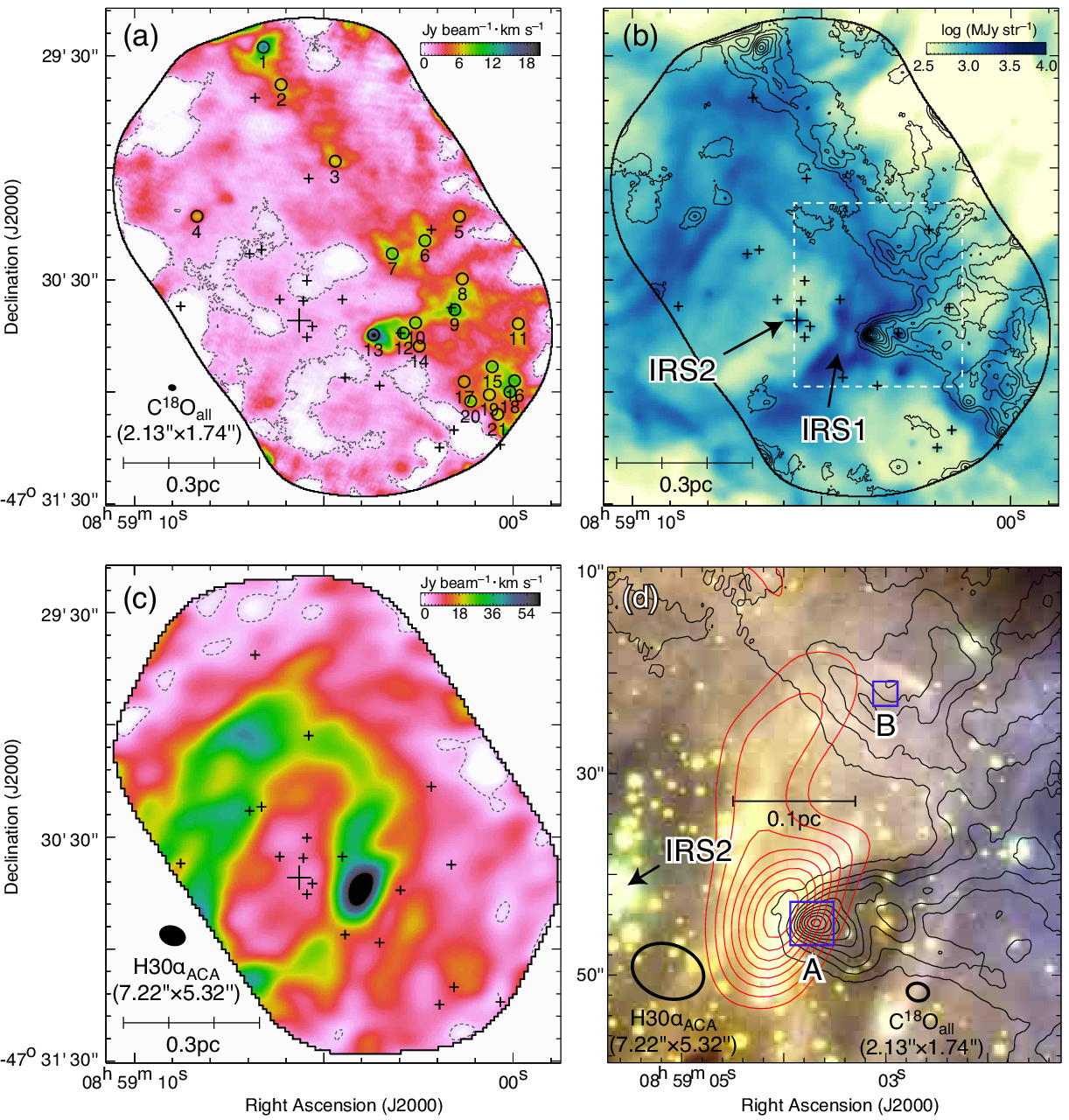}
 \end{center}
 \caption{(a) Intensity distributions of the C$^{18}$O$_{\rm all}$ data integrated over a velocity range of $-2$--$+12$\,km\,s$^{-1}$. Dashed contours indicate the zero level of the C$^{18}$O$_{\rm all}$ image. Circles indicate the positions of the C$^{18}$O condensations. (b) Contour map of the C$^{18}$O$_{\rm all}$ image in (a) superimposed on the {\it Spitzer}/IRAC 5.8\,$\mu$m image \citep{wol2008}. Contours start at 1\,Jy\,beam$^{-1}$\,km\,s$^{-1}$ with steps of 2\,Jy\,beam$^{-1}$\,km\,s$^{-1}$. 
 (c) H30$\alpha_{\rm ACA}$ image, with the dashed contours plotted at the zero level. The velocity integration range is from $-20$ to $30$\,km\,s$^{-1}$.
 In (a), (b) and (c), the large cross indicates the central O5.5 star IRS\,2, whereas small crosses indicate the O star candidates \citep{win2011}. 
 (d) A VLT near-infrared $J$-(blue), $H$-(green), and $K_{\rm s}$-(red) image \citep{wol2006} shown in the region indicated by the white box in (b). Black contours show the C$^{18}$O$_{\rm all}$ distribution plotted in (b), whereas H30$\alpha_{\rm ACA}$ is shown in the red contours starting at 10$\sigma$ (23\,Jy\,beam$^{-1}$\,km\,s$^{-1}$) with steps of 3$\sigma$ (6\,Jy\,beam$^{-1}$\,km\,s$^{-1}$). 
 The two boxes marked by blue lines indicate the regions of Source\,A and Source\,B shown in Figures\,\ref{233}(a) and (b), respectively.
 }\label{fig1}
\end{figure}

From the C$^{18}$O$_{\rm all}$ data with a resolution of $\sim$2$''$, Figures\,\ref{fig1}(a) and (b) show the C$^{18}$O intensity distribution integrated over $-2$ to $+14$\,km\,s$^{-1}$, which covers the velocity ranges of the ring cloud, finger cloud, and bridge feature. 
The majority of the emission arises from the ring cloud.
The C$^{18}$O$_{\rm all}$ map shows filamentary structures with widths of 0.1--0.2\,pc, extending along the northwest-southeast direction, whose rim at the bottom side is illuminated by nearby O stars, as indicated by the bright 5.8\,$\mu$m emission (Figure\,\ref{fig1}(b)).
The filamentary structures include many condensations.
The condensations were identified from the C$^{18}$O$_{\rm all}$ integrated intensity map in Figure\,\ref{fig1}(a) using the ``clumpfind'' algorithm \citep{clumpfind_wil1994}, where the lowest detection level was set to 5\,$\sigma$ ($=$\,3\,Jy\,beam$^{-1}$\,km\,s$^{-1}$). Distributions of the 21 identified condensations are indicated by the circles in Figure\,\ref{fig1}(a), whereas their physical properties are summarized in Table\,\ref{tab:1.5}.

The condensations have typical radius $r_{\rm C^{18}O}$ of 0.01--0.02\,pc (which correspond to $\sim$2000--4000\,AU) and are separated by 0.05--0.3\,pc with each others.
The H$_2$ column density $N_{\rm H_2}$ and H$_2$ mass $M_{\rm H_2}$ of the condensations were estimated assuming Local Thermodynamic Equilibrium (LTE).
These condensations show brightness temperatures $T_{\rm b}$ of 30--50\,K in the $^{13}$CO$_{\rm all}$ data, and these values were used for the excitation temperature $T_{\rm ex}$ in the calculations.
A [$^{16}$O]/[$^{18}$O] ratio of 550 and a [H$_2$]/[$^{12}$CO] ratio of $10^4$ was adopted (e.g., \cite{fre1982, leu1984, wil1994}).
The derived $N_{\rm H_2}$ are as high as an order of $10^{23}$--$10^{24}$\,cm$^{-2}$ at the peak positions of the condensations, whereas the $M_{\rm H_2}$ ranges from $\sim$10$M_\odot$ to $\sim$100\,$M_\odot$. We derived the virial mass, $M_{\rm vir}$ = 210$r_{\rm C^{18}O}$(2$\sqrt{\rm 2ln2}\sigma{_v}$)$^2$, assuming a uniform density inside the cores \citep{Mac1988} without effects of the magnetic field and external pressure. The virial parameter, $\alpha_{\rm vir}$ = $M_{\rm vir}$/$M_{\rm H_2}$, is the order of unity, suggesting that these condensations are gravitationally bounded.
There are several C$^{18}$O condensations with remarkably high $N_{\rm H_2}$ of $10^{24}$\,cm$^{-2}$, i.e., \#1, \#9, \#10, \#12, \#13, \#16, \#18, and \#19. 
The most dense and massive condensation \#13 is located immediately west of IRS\,1.

Figure\,\ref{fig1}(c) shows the ionized gas distribution in the H30$\alpha_{\rm ACA}$ image. It shows a ring-like distribution, surrounding IRS\,2, and is peaked at IRS\,1.
Figure\,\ref{fig1}(d) shows a magnified view of the C$^{18}$O$_{\rm all}$ and H30$\alpha_{\rm ACA}$ maps toward the condensation \#13 and IRS\,1, superimposed on the Very Large Telescope (VLT) near-infrared image \citep{wol2006}, indicating that the eastern rim of the condensation \#13 follows the western rim of the bright ridge of IRS\,1, and IRS\,1 is spatially coincident with the H30$\alpha_{\rm ACA}$ distribution. This result suggests that the bright infrared emission of IRS\,2 arises from the strong irradiation of the outer envelope of the condensation by the adjacent IRS\,2.

\subsection{Discovery of two compact mm-sources: Source\,A and Source\,B}\label{sec:srcAB}
\begin{figure}
 \begin{center}
  \includegraphics[width=15cm]{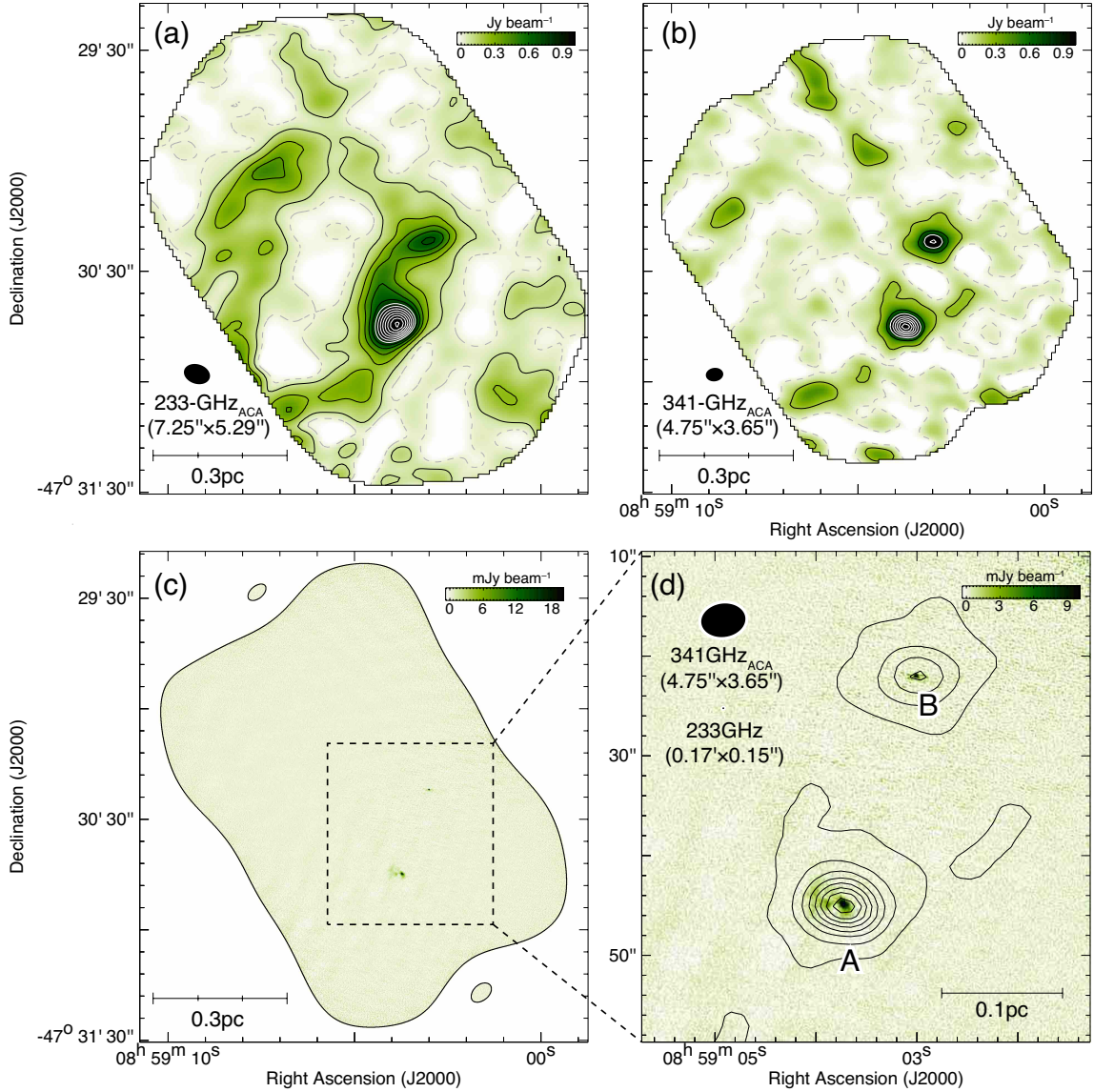}
 \end{center}
 \caption{
(a) A 233-GHz$_{\rm ACA}$ image in the same region in Figures\,\ref{fig1}(a)--(c), with contours starting at 5$\sigma$ (0.19\,Jy\,beam$^{-1}$) with steps of 10$\sigma$ (0.37\,Jy\,beam$^{-1}$). (b) A 345-GHz$_{\rm ACA}$ image, with contours starting at 5$\sigma$ (0.07\,Jy\,beam$^{-1}$) with steps of 10$\sigma$ (0.15\,Jy\,beam$^{-1}$). Contours with dashed lines are plotted at the zero level.
(c) A 233-GHz image.
(d) An enlarged image of the 233-GHz data toward the box in (c). Contours show the 341-GHz$_{\rm ACA}$ distribution, plotted at the same levels in (b).
 }\label{cont}
\end{figure}

Figures\,\ref{cont}(a) and (b) show radio continuum emission of 233-GHz$_{\rm ACA}$ and 341-GHz$_{\rm ACA}$, respectively.
Two bright sources are detected in these two images. Hereafter, the two sources are referred to ``Source\,A'' and ``Source\,B''.
Although the 341-GHz$_{\rm ACA}$ image presents several clumpy structures without extended emission, the 233-GHz$_{\rm ACA}$ image shows a ring-like extended emission, which resembles the recombination line emission in Figure\,\ref{fig1}(c). It is suggested that both the emission from cool dust grains and ionized gas are observed in the 233-GHz$_{\rm ACA}$ image.
Figure\,\ref{cont}(c) shows distributions of the cool dust emission in the 233-GHz image with the highest spatial resolution in this study.
Two sources, Source\,A and Source\,B, are significantly detected, without any significant detections (above 3$\sigma$) of other C$^{18}$O condensations in this region, indicating that there are no compact/developed structures, such as protostellar disks, inside them. Figure\,\ref{cont}(d) presents a magnified view of the 233-GHz image toward the corresponding region of Figure\,\ref{fig1}(d), with a contour map of 341-GHz$_{\rm ACA}$.

Source\,A coincides with the condensation \#13 west of IRS\,1, whereas Source\,B is located 0.1--0.2\,pc northwest of IRS\,1 without any counterparts in the C$^{18}$O condensations (Figure\,\ref{fig1}(c)). 
As Source\,B is observed in the ACA map (Figure\,\ref{fig1}(c)), which covers extended emission, the possibility of resolved-out can be dismissed to interpret the absence of C$^{18}$O condensation.

\begin{figure}
 \begin{center}
  \includegraphics[width=16cm]{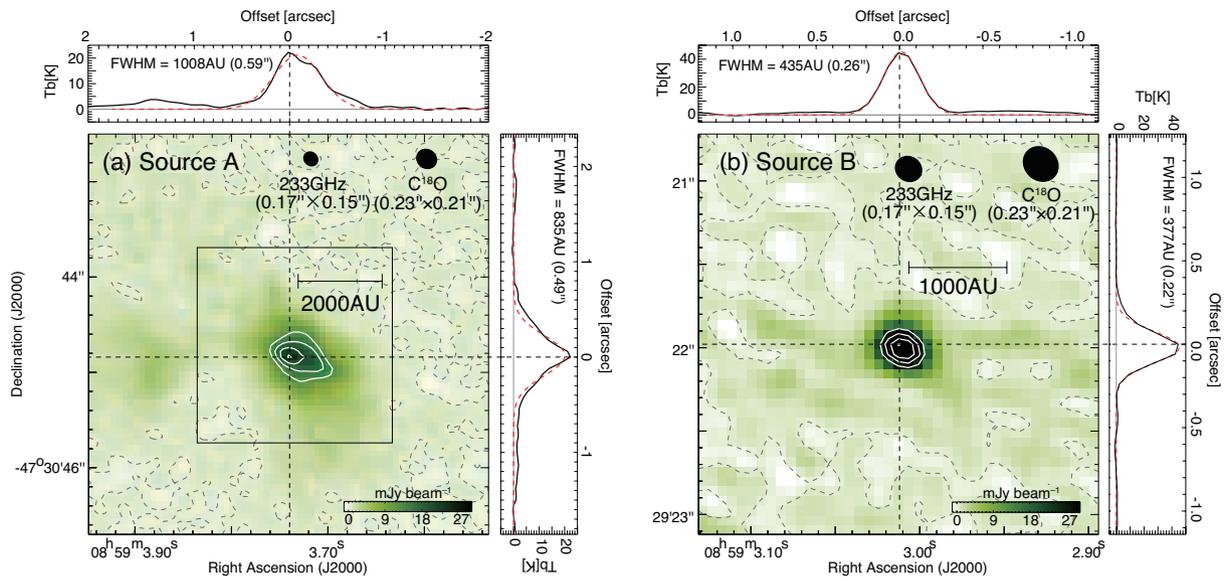}
 \end{center}
 \caption{(a, b) Magnified views of the 233-GHz image for Source\,A and Source\,B. The white contours in (a) and (b) start at half of the peak intensities, 24.9 and 48.8\,mJy\,beam$^{-1}$, with steps of 4 and 8\,mJy\,beam$^{-1}$, respectively. The dashed contours are plotted at the zero level. Intensity profiles along the dashed lines across the peaks of the 233-GHz emission are plotted in black lines at the upper- and right-sides of the panels in units of $T_{\rm b}$, with the best-fit Gaussian curves overlaid in dashed red lines.}\label{233}
\end{figure}

Figure\,\ref{233} shows an enclosed-view of Source\,A and Source\,B in the 233-GHz map with a resolution of $\sim0''.16$.
Both sources exhibit a single peak component.
Source\,A shows a parallelogram-like distribution with a peak flux density of 24.9\,mJy\,beam$^{-1}$, whereas Source\,B with a peak flux density of 48.8\,mJy\,beam$^{-1}$ has a circular distribution, which is not fully resolved at the present spatial resolution. 
The obtained peak brightness temperatures of Source\,A and Source\,B correspond to $T_{\rm b}$ of 22 and 42\,K, respectively. The relatively weak, diffuse emission located east of Source\,A, as shown in Figure\,\ref{233}(a), coincides with the western rim of IRS\,1, which may due to free-free emission, rather than thermal dust emission. We performed the aperture photometry toward both sources at the central 12$''$ diameter in the 233-GHz map. The total fluxes of Source\,A and B are 0.73 and 0.17\,Jy, respectively. These figures account for approximately 24\% and 13\% of the total fluxes measured with the 233-GHz$_{\rm ACA}$ map for the same areas.
This result indicates that the 233-GHz data obtained with the 12-m array overlooks a significant amount of flux from the extended dust emission toward Source\,A and Source\,B.

The VLT/ISACC and NACO $J$, $H$, and $K_{\rm s}$ images obtained by \citet{wol2006} and \citet{der2009} indicate no significant excess of emission toward Source\,A and Source\,B. This is also the case for the four {\it Spitzer}/IRAC bands (3.4, 4.5, 5.6, and 8.0\,$\mu$m; \cite{wol2006}). Unfortunately, there are no available data with high angular resolution at infrared wavelengths exceeding 8\,$\mu$m. Because of the non-detection of the available infrared observations, the two sources at an early phase of star formation in which the protostellar object driving the molecular outflows (Section\,3.2) are deeply embedded. This nature is consistent with the non-detection of H30$\alpha$ emission from the continuum peaks with the 12-m array, meaning that the young sources do not start to ionize their surroundings.

\subsection{Physical properties of Source\,A and Source\,B}
Tables\,\ref{tab:2.5} and \ref{tab:3} summarize the physical parameters of Source\,A and Source\,B estimated from the 233-GHz and 341-GHz$_{\rm ACA}$ maps. 

\subsubsection{341-GHz$_{\rm ACA}$ data}

The 233-GHz$_{\rm ACA}$ image recovers extended structures (Figure\,\ref{cont} (a)), but the ring-like morphology is quite similar to the H30$\alpha$ distribution (Figure\,\ref{fig1}(c)). This means that the ionized gas (i.e., free-free) emission contaminates the 233-GHz$_{\rm ACA}$ image (see Section~\ref{sec:srcAB}). We thus used the 341-GHz$_{\rm ACA}$ map (Figure\,\ref{cont} (b,d)), whose distribution reasonably agrees with that in C$^{18}$O (Figure\,\ref{fig1} (a)) tracing cold/dense materials, to estimate the core properties of Source A and B. The physical parameters measured from the 341-GHz$_{\rm ACA}$ map, which has an angular resolution of $\sim4'$, are as follows;
The radii of Source\,A and Source\,B at the half maxima, $r_{\rm dust}$, are as small as $\sim$0.02 and $\sim$0.03\,pc, respectively.

The $N_{\rm H_2}$ and $M_{\rm H_2}$ of the sources were calculated from the peak and total fluxes of the sources ($S_{\rm peak}$ and $S_{\rm int}$) at the optically-thin limit as follows;
\begin{eqnarray}
%\begin{displaymath}
N_{\rm H_2} \ &=& \ \frac{R_{\rm g/d} S_{\rm peak, \nu}}{2.8 m_{\rm H} \kappa_{\nu} B_{\nu} (T_{\rm dust}) \Omega_{\rm beam}} \label{eq:nh2}\\
M_{\rm H_2}  \ &=& \ \frac{R_{\rm g/d} S_{\rm int, \nu} D^2}{\kappa_{\nu} B_{\nu} (T_{\rm dust})}, \label{eq:mh2}
%\end{displaymath}
\end{eqnarray}
where $T_{\rm dust}$ is the dust temperature and $R_{\rm g/d}$ is the gas-to-dust ratio. 
Here, an $R_{\rm g/d}$ of 100 was assumed.
The dust opacity $\kappa_{\nu}$ at 341-GHz was assumed to be 1.8\,cm$^{-2}$\,g$^{-1}$ from a thin ice mantle model of \citet{oss1994} for a density of $10^6$\,cm$^{-3}$.
$B_{\nu} (T_{\rm d})$ is the Planck function for $T_{\rm d}$.
Far-infrared and submillimeter continuum observations indicate that cold dust components embedded within the H{\sc ii} regions typically have a $T_{\rm dust}$ of 15--30\,K (e.g., \cite{and2012, tor2017a, mot2018}). Considering the proximity of the exciting stars to Source\,A and Source\,B, which may lead to higher temperatures, $T_{\rm dust}$ values of 20--40\,K were assumed in this study.
As a result, the $N_{\rm H_2}$ values at the peaks of the two sources were calculated to be $1.1$--$2.7\times10^{24}$\,cm$^{-2}$ for Source\,A and $0.5$--$1.2\times10^{24}$\,cm$^{-23}$ for Source\,B, whereas the $M_{\rm H_2}$ values were as large as 32--81\,$M_\odot$ for Source\,A and 16--41\,$M_\odot$ for Source\,B. The number densities of the condensations $n_{\rm H_2}$ were calculated as an order of $10^7$\,cm$^{-3}$ from the $M_{\rm H_2}$ assuming a sphere.

The 3$\sigma$ mass-detection limit of the 341-GH$_{\rm ACA}$ image is approximately 3--6\,$M_{\odot}$, depending on the $T_{\rm d}$ assumption (15--30\,K). Although we detected some additional C$^{18}$O cores, such as \#4 and \#5, their mass is similar to or lower than the continuum detection limit.

\subsubsection{233-GHz data}
The physical parameters of Source\,A and Source\,B measured from the high-resolution 233-GHz data are as follows;
The $r_{\rm dust}$ values of Source\,A and Source\,B were measured by fitting the horizontal and vertical profiles across the peaks with a Gaussian function, with values of $\sim$460 and $\sim$200\,AU, respectively (Figure\,\ref{233}).
Because it is difficult to measure the optical depths of the sources with the present dataset, we assumed the optically-thin case to estimate the $N_{\rm H_2}$ and $M_{\rm H_2}$ using Equations (\ref{eq:nh2}) and (\ref{eq:mh2}) with a $\kappa_\nu$ of 0.899\,cm$^{-2}$\,g$^{-1}$, and an $R_{\rm g/d}$ of 100. Since the observed peak $T_{\rm b}$ values of 22\,K for Source\,A and 42\,K for Source\,B in the 233-GHz image are presumably the lower limit as the actual temperature, we adopted $T_{\rm d}$ = 100\,K based on the recent molecular line measurements toward the G353.273+0.641 high-mass protostellar disk \citep{mot2019}.
The derived $N_{\rm H_2}$ value is approximately $6\times10^{24}$\,cm$^{-2}$ for the two sources, with the $M_{\rm H_2}$ values calculated as $\sim$5\,$M_\odot$ for Source\,A and $\sim$2\,$M_\odot$ for Source\,B. 
Observations at lower frequencies are crucial for measuring $N_{\rm H_2}$ and $M_{\rm H_2}$ more accurately by quantifying the optical depth of the dust emission. 

Note that we could not detect any other emission in the 233-GHz continuum image (see Figure\,\ref{cont}). Recent ALMA surveys with the 12-m array in the nearby low-mass star-forming regions demonstrated that it is hard to detect millimeter/submillimeter continuum emission in starless sources due to the lack of compact features, such as protostellar disks, inside them (\cite{dun2016}; see also \cite{tok2020}). It is thus likely that the other cores in the observed RCW~38 field are intrinsically in the starless phase.

\subsubsection{Velocity distribution of the C$^{18}$O emission in Source\,A}
\begin{figure}
 \begin{center}
  \includegraphics[width=7cm]{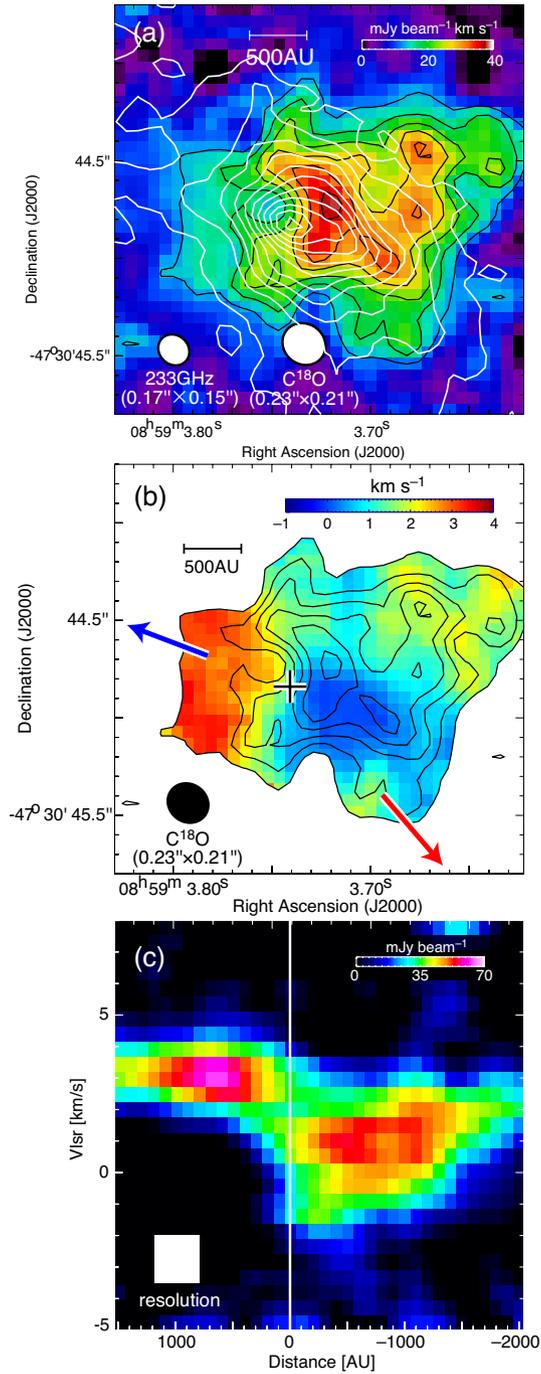}
 \end{center}
 \caption{(a) Intensity distribution of the C$^{18}$O data (color and black contours) and the 233-GHz map (white contour) for Source A.
The velocity integration range is $-2$ to $+4$\,km\,s$^{-1}$. The black contours start at 3$\sigma$ (135\,mJy\,beam$^{-1}$\,km\,s$^{-1}$) with steps of 1$\sigma$ (45\,mJy\,beam$^{-1}$\,km\,s$^{-1}$), whereas the white contours start at 3$\sigma$ (1.35\,mJy\,beam$^{-1}$) with steps of 5$\sigma$ (2.25\,mJy\,beam$^{-1}$).
(b) First moment map of the C$^{18}$O data (color) superimposed with the contour map of the C$^{18}$O data in (a). The cross indicates the peak position of the 233-GHz image. The two arrows indicate the directions of the blueshifted and redshifted outflow lobes shown in Figure\,\ref{outflowA}. (c) Position-velocity diagram of the C$^{18}$O data. 
the C$^{18}$O data were smoothed to be a velocity resolution of 1.5\,km\,s$^{-1}$.
The vertical white line indicates the direction of the 233-GHz peak of Source\,A.
%and the dashed lines show the Keplerian velocities calculated for a central mass of 1 (white), 3 (red), and 8\,$M_\odot$ (yellow).
}\label{c18o}
\end{figure}

The high resolution C$^{18}$O data were analyzed to investigate the molecular gas components associated with Source\,A.
Figure\,\ref{c18o}(a) shows the intensity distribution of the C$^{18}$O data integrated over a velocity range of $-2$--$+4$\,km\,s$^{-1}$, in which the C$^{18}$O emission was detected in the presented region. 
The C$^{18}$O distribution is more extended than the 233-GHz distribution, and the peak position does not perfectly coincide with the peak of the 233-GHz map, as shown by the white contours.

Figure\,\ref{c18o}(b) shows a first moment map of the C$^{18}$O data. 
A clear velocity gradient along the southwest-northeast direction is present, which can also be observed in the position-velocity (p-v) diagram in Figure\,\ref{c18o}(c).
The p-v diagram shows two velocity components at $\sim-1$ and $\sim3$\,km\,s$^{-1}$ located west and east to the 233-GHz peak, respectively, and the 233-GHz peak is located in the center of the intermediate velocity feature connecting the two velocity components. 
These observed signatures suggest rotational motion of the C$^{18}$O component centered on the 233-GHz peak. 
%If we assume the pure Keplrian velocities ($\propto r^{-0.5}$) whose inclination is edge-on, the mass of the central object is likely $\sim$1--3\,$M_{\odot}$. 
However, the directions of the outflow lobes which will be presented in the next subsection are not perpendicular to the direction of the velocity gradient (see arrows in Figure\,\ref{c18o}(b)). 
This is inconsistent with the hypothesis that the observed C$^{18}$O component is a protostellar disk.
Infall motion of gas, whose velocity is proportional to $r^{-1}$ (e.g., \cite{oha2014}), is an alternative idea to explain the observed velocity gradient. 
It is difficult to distinguish these two possibilities with the sensitivity and velocity resolution of the observed p-v diagram.
If the velocity gradient was purely attributed to the Keplerian rotation, as the C$^{18}$O distribution does not perfectly match the 233-GHz distribution, the obtained result does not immediately indicate that Source\,A harbors a protostar with a mass of a few $\times$\,$M_\odot$.
Indeed, \citet{zha2019} highlighted the importance of the choice of molecular lines in investigating the rotational motion to directly determine the mass of the central object.

\subsection{High-velocity molecular outflows for Source\,A and Source\,B}\label{sec:outflow}
\begin{figure}
 \begin{center}
  \includegraphics[width=6cm]{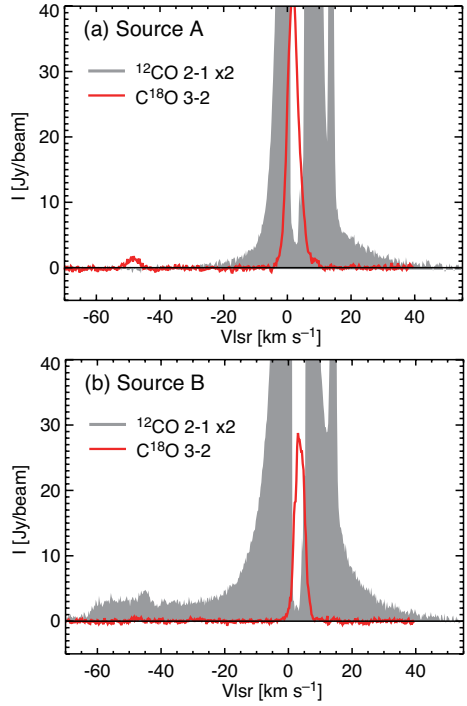}
 \end{center}
 \caption{Spectra of the $^{12}$CO$_{\rm ACA}$ and C$^{18}$O$_{\rm ACA}$ data near the peaks of the 233-GHz map for Source A and Source B. The intensities of the $^{12}$CO$_{\rm ACA}$ are doubled.} 
\label{spec}
\end{figure}

In this subsection we report the discovery of high-velocity molecular outflows associated with Source\,A and Source\,B.
The physical properties of the outflow lobes are summarized in Table\,\ref{tab:4}.
Figure\,\ref{spec} shows the spectra obtained from the $^{12}$CO$_{\rm ACA}$ and C$^{18}$O$_{\rm ACA}$ toward the 341-GHz$_{\rm ACA}$ peaks of Source\,A and Source\,B (Figure\,\ref{fig1}(d)).
Wing features with velocity widths of $\sim$30--70\,km\,s$^{-1}$ are clearly present in the $^{12}$CO spectra for both the blueshifted and redshifted velocity ranges of Source\,A and Source\,B.

\begin{figure}
 \begin{center}
  \includegraphics[width=13cm]{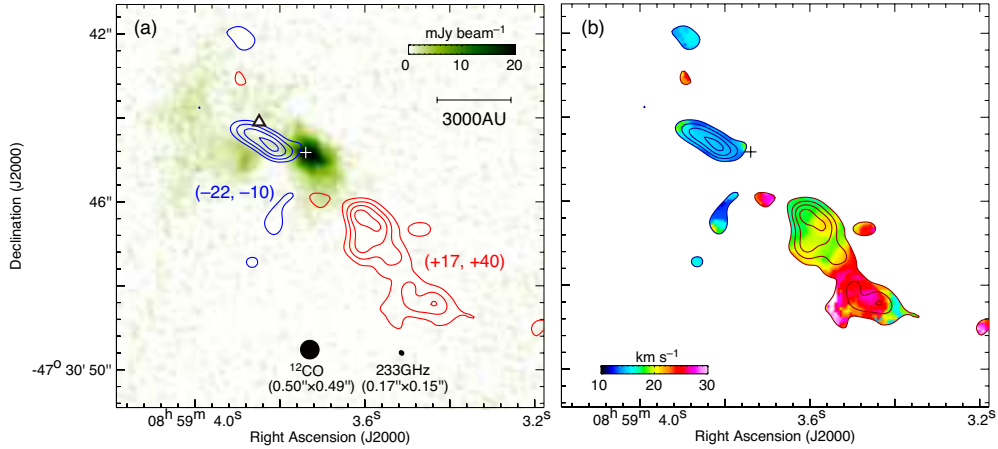}
 \end{center}
 \caption{(a) $^{12}$CO intensity distributions of the blueshifted and redshifted outflow lobes shown in blue and red contours, respectively, with the 233-GHz background image. The velocity ranges of the lobes are shown in parentheses in units of km\,s$^{-1}$. The contours stars at 3\,$\sigma$ with steps of 1.5\,$\sigma$, where the 1\,$\sigma$ value of the blueshifted and redshifted maps of Source\,A are 30 and 65\,mJy\,beam$^{-1}$\,km\,s$^{-1}$, respectively. The triangle depicts the approximate position of the vibrational transition of the H$_2$ emission detected by VLT \citep{der2009}. (b) Color map of $|v_{\rm mom1} - v_{\rm sys}|$ for the lobes, with the $^{12}$CO contour maps plotted in (a), where $v_{\rm mom1}$ is the first moment generated from the $^{12}$CO data for the velocity ranges shown in panel (a), whereas the systemic velocity of the source $v_{\rm sys}$ was measured from the C$^{18}$O$_{\rm ACA}$ data as 2\,km\,s$^{-1}$ (Figure\,\ref{spec}). The cross indicates the 233-GHz peak for Source\,A. }\label{outflowA}
\end{figure}

Figure\,\ref{outflowA}(a) shows the spatial distributions of the outflow lobes in Source\,A from the high-resolution $^{12}$CO data, whereas Figure\,\ref{outflowA}(b) presents velocity distributions of the outflow lobes, defined as $|v_{\rm mom1} - v_{\rm sys}|$, where $v_{\rm mom1}$ and $v_{\rm sys}$ are the first moment and systemic velocity of the source, respectively.
The $v_{\rm sys}$ value was determined from the C$^{18}$O$_{\rm ACA}$ spectra in Figure\,\ref{spec} as 2\,km\,s$^{-1}$.

As shown in Figure\,\ref{outflowA}(a), the blueshifted and redshifted lobes are elongated from Source\,A toward the northeast-southwest direction by $\sim$3200 and $\sim$9100\,AU, respectively.
These two lobes are not perfectly aligned with each other, and the outflow axis does not correspond to the rotational axis of the C$^{18}$O component (see arrows in Figure\,\ref{c18o}(b)).
In the blueshifted velocity range, other weak emission is observed north and south of the lobe, but it is difficult to determine whether these features are associated with Source\,A, as the signal-to-noise (S/N) ratio of the present ALMA molecular line data is not high.
The triangle in Figure\,\ref{outflowA}(a) depicts the position of the H$_2$ emission at 2.12\,$\mu$m reported by \citet{der2009} using VLT/NACO, which is located near the tip of the blueshifted lobe, suggesting the presence of shocked molecular gas caused by the outflow.

As shown in Figure\,\ref{outflowA}(b), the redshifted lobe has a velocity gradient along the lobe, whereas such a gradient is not clear in the blueshifted lobe.
Given the maximum velocities $v_{\rm max}$ of  27 and 38\,km\,s$^{-1}$, the dynamical timescales of the blueshifted and redshifted lobes $t_{\rm dyn}$ were estimated as 560 and 1100\,yr, respectively, depending on their inclination angles, $i$, which is defined as 0$^{\circ}$ for the plane of the sky. It is not simple to measure the H$_2$ masses of the outflow lobe $M_{\rm lobe}$, because this value strongly depends on the line opacity, excitation temperature, inclination, etc; moreover it is difficult to distinguish the effects of these parameters for the relatively low-S/N ratio of the line data obtained with the 12-m array.
However, using the synthetic observation technique, \citet{off2011} demonstrated that the CO-to-H$_2$ conversion factor can be used to track the actual mass of the outflow over different epochs, although the derived $M_{\rm lobe}$ is generally a factor of 5--10 smaller than the actual mass due to the opacity effect.
Therefore, in this study we adopt a CO-to-H$_2$ conversion factor of $2\times10^{20}$\,(K\,km\,s$^{-1}$)$^{-1}$\,cm$^{-2}$ \citep{bol2013}, as a conservative way to estimate the lower-limit of  $M_{\rm lobe}$.
The derived $M_{\rm lobe}$ values of the blueshifted and redshifted lobes were 0.06 and 0.4\,$M_\odot$, respectively (Table\,\ref{tab:4}). If we assume the inclination angle of 30$^{\circ}$--70$^{\circ}$, the mass outflow rates $\dot{M_{\rm lobe}}$ of the two lobes were derived as (0.4--1.6)\,$\times$10$^{-4}$ and (1.3--6.3)$\,\times$10$^{-4}$\,$M_\odot$\,yr$^{-1}$, respectively.

\begin{figure}
 \begin{center}
  \includegraphics[width=13cm]{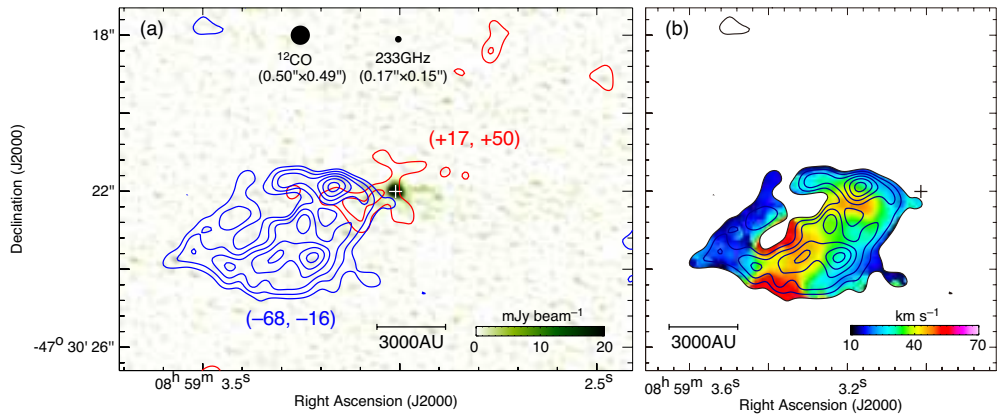}
 \end{center}
 \caption{The same plots shown in Figure\,\ref{outflowA} are presented here, but for Source\,B. In (a), the 1\,$\sigma$ levels of the $^{12}$CO maps are 90 and 100\,mJy\,beam$^{-1}$\,km\,s$^{-1}$ for the blueshifted and redshifted lobes, respectively. In (b), only the blueshifted lobe is plotted. }\label{outflowB}
\end{figure}

Compared with the lobes in Source\,A, the blueshifted lobe in Source\,B is relatively expanded, showing a cocoon-like shape. 
The $v_{\rm max}$ of the blushifted lobe is found to be as high as 67\,km\,s$^{-1}$, and the velocity increases within the cocoon (Figure\,\ref{outflowB}(b)).
In contrast, the redshifted lobe does not show a clear shape; rather a weak, diffuse component is observed toward Source\,A. 
Although two other weak features are observed northwest of Source\,B in teh redshifted velocity range, the association of these features with the redshifted lobe cannot be confirmed at this time.
The projected length of the blueshifted lobe is $\sim10000$\,AU, providing a short $t_{\rm dyn}$ of 750\,yr (Table\,\ref{tab:4}). The possible range of the mass-loss rate is (0.7--3)\,$\times$\,10$^{-3}$\,$M_{\odot}$\,yr$^{-1}$ with the same assumption of $i$ in the previous paragraph.

\section{Discussion}
\subsection{Source\,A and Source\,B; nurseries of massive protostars?}\label{dis:AB}
The present ALMA data revealed two compact mm-sources at the center of the young super star cluster RCW\,38.
The two sources are embedded within dust condensations with $r_{\rm dust}$ of $\sim$0.02--0.03\,pc and $M_{\rm H_2}$ greater than $\sim$20--30\,$M_\odot$ (Table\,\ref{tab:3}). These condensations seem to satisfy the initial conditions for the massive star formation (e.g., \cite{tan2014}), at least in their mass-concentrated nature. The mechanical force of the outflows, $F_{\rm out}$ = $M_{\rm lobe}v_{\rm max}$/$t_{\rm dyn}$, is the order of 10$^{-2}$\,$M_{\odot}$\,km\,s$^{-1}$\,yr$^{-1}$, which is consistent with that of known high-mass protostars compiled by \citet{mau2015a}. Since the mass accretion rate $\dot{M_{\rm acc}}$ correlates with the outflow mass rate ($\dot{M_{\rm acc}} \simeq\ $a few$\ \times\ \dot{M_{\rm lobe}}$) \citep{beu2002, mau2015a}, the observed high $\dot{M_{\rm lobe}}$ values suggest a high $\dot{M_{\rm acc}}$ on the order of $10^{-4}$--$10^{-3}$\,$M_\odot$\,yr$^{-1}$ for Source\,A and Source\,B, which is consistent with the values obtained with massive star formation models (e.g., \cite{mck2003, hos2009, hos2010}).
These observational signatures imply that Source\,A and Source\,B harbor massive protostars that can drive energetic outflows. Note that the two sources exhibit a single peak at the 270\,AU resolution, indicating no signs of fragmentation.

Further, the short $t_{\rm dyn}$ values of the outflow lobes in Source\,A and Source\,B suggest that these protostars are in an early evolutionary stage.
If $t_{\rm dyn}$ is interpreted as the age of the protostar driving the outflow, it should coincide with the accretion time $t_{\rm acc}$, suggesting that the total $M_{\rm lobe}$ of the outflow lobes, $\sim$0.5\,$M_\odot$ for Source\,A and $\sim$1.6\,$M_\odot$ for Source\,B.%, would be similar to the current masses of the protostars.
%It is quite important to constrain the masses of the protostars in Source\,A and Source\,B based on the high-resolution and high-sensitivity molecular line data in order to understand the evolutionary stage of the protostars.
%Although the present ALMA C$^{18}$O data presents a velocity gradient that can be possibly interpreted as a Keplerian rotation, which suggests a mass of the central object to be a few $M_\odot$, the sensitivity and velocity resolution of the ALMA data are not enough to distinguish from the infall motion of gas, as mentioned in Section\,3.3.3.

\if0
If it is assumed that the velocity gradient observed in Source\,A is attributed to the Keplerian rotation of the protostellar disk, the mass of the protostar in Source\,A can be calculated as a few $M_\odot$ (Figure\,\ref{c18o}(b)), which is consistent with the mass predicted from the $t_{\rm dyn}$ and $\dot{M_{\rm lobe}}$ of the outflow lobes.
Because the C$^{18}$O emission is absent in Source\,B (Figure\,\ref{fig1}(c)), it is difficult to estimate the mass of the protostar in Source\,B from the gas motion.
However, because Source\,B is smaller than Source\,A (Figure\,\ref{233} and Table\,\ref{tab:3}), it is reasonable to assume that the stellar mass in Source\,B is similar with that of Source\,A.
\fi

\subsection{Infrared-quiet sources}
%As the large $\dot{M_{\rm lobe}}$ values imply large $\dot{M_{\rm acc}}$ for Source\,A and Source\,B, and considering the small $t_{\rm dyn}$, the protostellar objects in Source\,A and Source\,B could be massive stars in the creation stage.
%Our ALMA data indicates that a large amount of gas ($>$3--13\,$M_\odot$) is present within a few hundred AU of Source\,A and Source\,B, and it can be expected that the protostellar objects will continue evolving by accreting the surrounding media.
Measuring the luminosity at infrared wavelength is another way to constrain the evolutionary stage of the protostars.
However, infrared images above 8.0$\mu$m are not available in RCW\,38 at an angular resolution of a few arcsecond.
Further, no significant emission was detected for Source\,A and Source\,B in the near- and mid-infrared wavelengths below 8.0\,$\mu$m; The VLT and {\it Spitzer}/IRAC images do not show any excess toward these two sources. If we adopt the accretion rate of 10$^{-4}$--10$^{-3}$\,$M_\odot$\,yr$^{-1}$ (Sect.\,\ref{dis:AB}) onto the central protostars inside Source\,A/B, Equation (22) in \citet{hos2009} tells us that the temperature of the accretion gas is 100--500\,K. In this case, the peak of the spectral energy distribution should be in near- or mid-infrared wavlength. Here, the {\it Spitzer}/IRAC 3.6, 4.5, 5.6, and 8.0\,$\mu$m archival images \citep{wol2006} were used to estimate the upper-limit of the luminosity at 3.6--8.0\,$\mu$m ($L_{3.6-8.0}$) for Source\,A and Source\,B, by integrating the upper-limits of the flux in the four bands.
The upper-limit of the flux in each band was measured as the standard deviation of the flux density distribution around a 4$''$ area of the source, and the derived upper-limits range $\sim$100--400\,MJy\,str$^{-1}$ for the four bands.
The $L_{3.6-8.0}$ values, which is not the bolometric luminosity, were then calculated as small as 3.0\,$L_\odot$ for Source\,A and 1.6\,$L_\odot$ for Source\,B.
Inclusion of the $J$, $H$, and $K_{\rm s}$ bands does not strongly change these values.

In these wavelengths, accreting gas as well as the protostar itself should be the major sources of the emission.
%If the accreting gas has a temperature of a few hundred Kelvin, the flux densities should increase in the near- and mid-infrared bands.
The accretion luminosity $L_{\rm acc}$ can be calculated as follows;
\begin{eqnarray}
%\begin{displaymath}
L_{\rm acc} \ &=& \ \frac{GM_{*}\dot{M_{\rm acc}}}{R_{*}} \ \simeq \ 1219 \left(\frac{M_{*}}{1\,M_\odot}\right) \times \left(\frac{\dot{M_{\rm acc}}}{10^{-3}\,M_\odot\,{\rm yr}^{-1}}\right) \times \left(\frac{26 R_\odot}{R_{*}}\right) \ [L_\odot], \label{eq:Lacc}
%\end{displaymath}
\end{eqnarray}
where $G$ is the gravitational constant, and $M_{*}$ is the stellar mass.
The stellar radius $R_{*}$ can be given as a function of $M_{*}$ and $\dot{M_{\rm acc}}$ \citep{sta1986,hos2009}:
\begin{eqnarray}
%\begin{displaymath}
R_{*} \ \simeq \ 26 \left(\frac{M_{*}}{1\,M_\odot}\right)^{0.27} \times \left(\frac{\dot{M_{\rm acc}}}{10^{-3}\,M_\odot\,{\rm yr}^{-1}}\right)^{0.41} \ [R_\odot]. \label{eq:rstar}
\end{eqnarray}
Assuming that $\dot{M_{\rm acc}}$ is equal to the total $\dot{M_{\rm lobe}}$ of the redshifted and blueshifted lobes, the expected $L_{\rm acc}$ for $M_{*}$ of 0.1 and 1\,$M_\odot$ is calculated as 140 and 770\,$L_\odot$ for Source\,A and 310 and 1700\,$L_\odot$ for Source\,B, respectively.
Even for a small $M_{*}$ of 0.1\,$M_\odot$, the derived $L_{\rm acc}$ is approximately two order of magnitude larger than the derived upper-limits of $L_{3.6-8.0}$ for Source\,A and Source\,B.

A reasonable interpretation of the low $L_{3.6-8.0}$ is that the dusty envelopes of Source\,A and Source\,B veil the luminous accreting gas in near- and mid-infrared wavelengths.
The two dust condensations have $N_{\rm H_2}$ of higher than $10^{24}$\,cm$^{-2}$ (see Figure\,\ref{fig1}(d) and Table\,\ref{tab:3}).
If dust opacity model in \citet{oss1994} is assumed, the optical depths of the dust emission at 3.6--8.0\,$\mu$m can be calculated to be $\sim$50--100 at $N_{\rm H_2} = 10^{24}$\,cm$^{-2}$. The emission of the accreting gas cannot be observed at these high optical depths.
However, the opacity effect strongly depends on the three-dimensional inner-structures of the sources.
If there is a hole created by the outflows in the condensation and if the inclination of the hole is not perpendicular to us, the emission from the accreting gas could be detected through the hole (e.g., \cite{mot2017,mot2019}).

Another idea is ``episodic accretion''.
This idea was originally studied in low-mass star formation, followed by recent studies on the massive star formation models (e.g., \cite{mey2017, mey2018, hos2016}).
In the episodic accretion scenario, protostars spend most of their time in the quiescent phase with a low $\dot{M_{\rm acc}}$, interspersed with short but intense accretion bursts caused by disc gravitational fragmentation followed by rapid migration of the fragments onto the protostar (e.g., \cite{dun2012, hos2016}).
Therefore, the quiescent phase with a low $L_{\rm acc} (\propto \dot{M_{\rm acc}}$) has a high probability of detection, whereas the $\dot{M_{\rm lobe}}$ of the outflows may be high, as it is calculated as a time-averaged value. 

Further investigation of these scenarios for Source\,A and Source\,B should be performed based on high-resolution observations.
Observations at lower frequencies, where dust opacity is less influential, for instance, for ALMA band\,3, are important accurately measuring the source masses.
A high angular resolution, increased by a factor of two or three, is crucial to obtain detailed mass distributions of the sources.
Observations with various molecular lines are also important for investigating the rotatinal motion of Source\,A and Source\,B. 
If Keplerian motion can be confirmed, the mass of the protostellar object can be measured.
Molecular line observations are also important to investigate the possibility of CO depletion in Source\,B.

\subsection{Formation of massive condensations in RCW\,38}
Source\,A and Source\,B are found inside molecular condensations embedded within the filamentary structures, with a width of $\sim$0.1--0.2\,pc, extending along the northeast-southwest direction in RCW\,38 (Figure\,\ref{fig1}).
The filamentary structures include twenty-one C$^{18}$O condensations with large $N_{\rm H_2}$ of $>10^{23}$\,cm$^{-2}$ and $M_{\rm H_2}$ of $>10$\,$M_\odot$ (Table\,\ref{tab:1.5}).
Although no 233-GHz continuum sources were detected in all the condensations except for \#13, which harbors Source\,A, there are no such massive/dense starless objects in nearby low-mass star-forming regions (e.g., \cite{oni2002,tac2002}). The observed core properties, especially mass and size, in RCW~38 are consistent with those in infrared dark clouds on the Galactic plane (e.g., \cite{oha2016}), indicating that the identified condensations are likely precursors of high-mass or at least intermediate-mass stars. Recent numerical simulations successfully demonstrated that a massive/compact dense core, whose mass and size are 100\,$M_{\odot}$ and 0.1\,pc, respectively, monolithically collapses into high-mass stars in a similar manner to low-mass star formation, which is referred to as the turbulent core accretion scenario (e.g., \cite{mck2003,kur2009}). \citet{kur2012} simulated the evolution of a larger system with a gas mass of $\sim$1000\,$M_{\odot}$ and $\sim$0.4\,pc, and pointed out that supersonic gas flows within the parental cloud produce filamentary structures with a few overdense spots, each of which looks like the turbulent core posted by \citet{mck2003}. One of the crucial questions is what mechanism drives the creation of such a massive (100--1000\,$M_{\odot}$) and compact gas clump with a size scale of sub-pc, adopted as the initial condition of the cloud-collapse simulations, without a significant mass consumption due to lower-mass star formation (see for a review \cite{fuk2021b}). 

The formation of the filamentary structures and massive condensations may be attributed to strong compression caused by feedback from adjacent O stars through a process known as ``collect and collapse'' \citep{elm1977}.
In this model, an expanding H{\sc ii} region sweeps the surrounding medium, forming a dense molecular shell, which is followed by the fragmentation and formation of new stars.
In the case of RCW\,38, however, the distance between the massive condensations and exciting source (IRS\,2) is too small for the accumulated gas to induce fragmentation. The lengths of semi-major and minor axis of the ring-like structure in H30$\alpha$ are $\sim$0.3\,pc, and $\sim$0.2\,pc, respectively (see Figure\,\ref{fig1}(c)). If we assume that a spherical geometry of the ionized gas and the location of IRS~2 is the center, the distance between the central source and Source\,A/B is at most $\sim$0.3\,pc even when we consider the projection effect. Because the small distance is roughly the same as the width of the filamentary structures, the gas density $n_{\rm H_2}$ in the accumulated layer can increase by a factor of only $\sim$2 from the initial gas density, if a plane-parallel is assumed; however, the model calculation requires that $n_{\rm H_2}$ be increased by one or two orders of magnitude (e.g., \cite{hos2006}).
Thus, it is more likely that the seeds of the massive condensations holding Source\,A and Source\,B were formed prior to the onset of ionization of the O stars, although it is still possible that the feedback from nearby O stars enhanced the accretion of the condensations (e.g., \cite{shima2017}).

On the other hand, as summarized in the Introduction, it is suggested that the formation of massive stars in RCW\,38 was triggered by a collision between two molecular clouds with a velocity separation of $\sim$10\,km\,s$^{-1}$ \citep{fuk2016}.
Recent ALMA observations of giant molecular clouds in the Large Magellanic Cloud found hub-filamentary structures \citep{mye2009}, in which massive stars are being formed, and these structures were likely formed by a large-scale colliding flow (\cite{fuk2019_N159, tok2019a}; see also \cite{and2016} for the case of NGC~6334 in the Milky Way). 
Numerical calculations indicate that such a supersonic collision would create a network of filaments with widths of a few 0.1\,pc at the interface of the collision \citep{ino2013, ino2018}.
\citet{fuk2021_cmf} analyzed the data of the numerical simulations of cloud-cloud collision in \citet{ino2013}, finding that the massive cores are efficiently formed within the filaments created through the cloud-cloud collision. These massive cores are separated with each others by $\sim$0.05--0.2\,pc, which are consistent with the separations of the observed condensations in RCW\,38 (Figure\,\ref{fig1}(a)). Although the thermal jeans instability with a volume density of 10$^4$--10$^5$\,cm$^{-3}$ may also describe the core separation, the observed density and mass are one or two orders of magnitude higher than those in the fragmentation mode. The observed massive/dense nature of the condensations along the filament in RCW\,38 could be indirect evidence that they may be products of an intense gas compression event, such as cloud-cloud collision.
Once such a compression makes gravitationally unstable massive cores, they eventually collapse into high-mass protostars following the core accretion scenario (e.g., \cite{tan2014}).

The competitive accretion (e.g., \cite{bon2001}) may alternatively realize the mass accretion rate of $\sim$10$^{-4}$\,$M_{\odot}$\,yr$^{-1}$ \citep{wan2010}, and thus we cannot fully exclude the competitive accretion scenario for high-mass star formation in the RCW~38 region. \citet{Ass2018} summarized that cloud-cloud collisions are one of the promising origins as a trigger of the formation of an isolated O-star system, such as M20 and RCW~120 \citep{tor2011,tor2015}, while the competitive accretion scenario may not be applicable to such a compact system of small mass \citep{fuk2021b}. Cloud-cloud collisions are a versatile mechanism which explains the high-mass star (cluster) formation throughout various interstellar scales. To quantitatively investigate whether or not the competitive accretion can be applied in RCW 38, we need a further detailed comparison between the observations and numerical models, which is beyond the scope of the present study.

\section{Summary}

The conclusions of the present study are summarized as follows.
\begin{enumerate}
\item The molecular gas associated with the Galactic super star cluster RCW\,38 is resolved at the 270-AU scale using the new ALMA cycle-3 band-6 and band-7 data. The C$^{18}$O image was used to identify the filamentary structures with a width of 0.1--0.2\,pc, which contains 21 condensations with $r_{\rm C^{18}O}$ of 0.01--0.02\,pc, peak $N_{\rm H_2}$ of $10^{23}$--$10^{24}$\,cm$^{-2}$, and $M_{\rm H_2}$ of 10--130\,$M_\odot$.
\item In addition, the high-resolution 233-GHz continuum data revealed two dense, compact mm-sources (Source\,A and Source\,B) with radii of $\sim$460 and $\sim$200\,AU. The $N_{\rm H_2}$ of the two sources was derived as an order of 10$^{24}$\,cm$^{-2}$, whereas the $M_{\rm H_2}$ were estimated as 13\,$M_\odot$ for Source\,A and 3\,$M_\odot$ for Source\,B at the optically-thin limit. 
Source\,A is coincident with one of the C$^{18}$O condensations, whereas no counterpart is found for Source\,B. 
The C$^{18}$O condensation holding Source\,A is located immediately west of IRS\,1, suggesting that the bright infrared ridge of IRS\,1 arises from irradiation of the eastern side of the condensation surface by IRS\,2.
\item The high-resolution C$^{18}$O image shows a velocity gradient of gas toward Source\,A, which can be interpreted as the rotation, although there is no robust evidence for directly tracing the Keplerian motion of Source\,A. Infall motion of gas is an alternative idea to interpret the velocity gradient. %High resolution and high sensitivity data are needed for further investigation.
\item The ALMA $^{12}$CO data detected molecular outflows associated with Source\,A and Source\,B. The outflows have high velocities of $\sim$30--70\,km\,s$^{-1}$ and large mass outflow rate of $\sim10^{-4}$\,$M_\odot$\,yr$^{-1}$ for Source\,A and $\sim10^{-3}$\,$M_\odot$\,yr$^{-1}$ for Source\,B. The derived values correspond to those expected for massive star formation models.
The short dynamical timescales of the outflows, less than 1000\,yr, suggest that the protostars in Source\,A and Source\,B may be massive stars in the creation stage.
\item Despite of the large $\dot{M_{\rm lobe}}$ which imply large $\dot{M_{\rm acc}}$ for Source\,A and Source, no significant excess was observed in the two sources at the near- and mid-infrared wavelengths. One idea is attenuation of the emission by dust in Source\,A and Source\,B. Another idea is the episodic accretion scenario, in which protostars spend most of their time in the quiescent phase, interspersed with short accretion bursts. We discussed the formation scenario of the filamentary structures and massive condensations in RCW~38, and eliminated the possibility of gas compression by the feedback from nearby O stars.
A collision between two molecular clouds may have triggered the formation of the filamentary structures in which the massive condensations were formed.
\end{enumerate}

\begin{ack}
This paper makes use of the following ALMA data: ADS/ JAO.ALMA\#2015.1.01134.S. ALMA is a partnership of ESO (rep- resenting its member states), NSF (USA) and NINS (Japan), together with NRC (Canada), MOST and ASIAA (Taiwan), and KASI (Republic of Korea), in cooperation with the Republic of Chile. The Joint ALMA Observatory is operated by ESO, AUI/NRAO and NAOJ. The authors thank Dr. Takashi Hosokawa for a valuable discussion.
This work was financially supported by Grants-in-Aid for Scientific Research (KAKENHI) of the Japanese society for the Promotion of Science (JSPS; grant numbers 18H05440, 18K13582, 15K17607, 15H05694, 24224005, and 26247026), and NAOJ ALMA Scientific Research Grant Number 2016-03B. We thank the anonymous referee for helpful comments, which significantly improved the manuscript.
\end{ack}

%\bibliographystyle{aasjournal_pasj} 
%\bibliography{reference}

\begin{thebibliography}{}
\expandafter\ifx\csname natexlab\endcsname\relax\def\natexlab#1{#1}\fi

\bibitem[{{Anderson} {et~al.}(2012){Anderson}, {Zavagno}, {Deharveng},
  {Abergel}, {Motte}, {Andr{\'e}}, {Bernard}, {Bontemps}, {Hennemann}, \&
  {Hill}}]{and2012}
{Anderson}, L.~D., {et~al.} 2012, \aap, 542, A10

\bibitem[{{Andr{\'e}} {et~al.}(2016){Andr{\'e}}, {Rev{\'e}ret}, {K{\"o}nyves},
  {Arzoumanian}, {Tig{\'e}}, {Gallais}, {Roussel}, {Le Pennec}, {Rodriguez}, \&
  {Doumayrou}}]{and2016}
{Andr{\'e}}, P., {et~al.} 2016, \aap, 592, A54

\bibitem[Ascenso(2018)]{Ass2018} Ascenso, J. 2018, The Birth of Star Clusters, Astrophysics and Space Science Library, Vol. 4, 424 (Cham: Springer), 1

\bibitem[{{Beuther} {et~al.}(2002){Beuther}, {Schilke}, {Sridharan}, {Menten},
  {Walmsley}, \& {Wyrowski}}]{beu2002}
{Beuther}, H., {Schilke}, P., {Sridharan}, T.~K., {Menten}, K.~M., {Walmsley},
  C.~M., \& {Wyrowski}, F. 2002, \aap, 383, 892

\bibitem[{{Beuther} {et~al.}(2017){Beuther}, {Walsh}, {Johnston}, {Henning},
  {Kuiper}, {Longmore}, \& {Walmsley}}]{beu2017}
{Beuther}, H., {Walsh}, A.~J., {Johnston}, K.~G., {Henning}, T., {Kuiper}, R.,
  {Longmore}, S.~N., \& {Walmsley}, C.~M. 2017, \aap, 603, A10

\bibitem[{{Bolatto}, {Wolfire}, and {Leroy}(2013){Bolatto}, {Wolfire}, \&
  {Leroy}}]{bol2013}
{Bolatto}, A.~D., {Wolfire}, M., \& {Leroy}, A.~K. 2013, \araa, 51, 207

\bibitem[Bonnell et al.(2001)]{bon2001} Bonnell, I.~A., Bate, M.~R., Clarke, C.~J., et al.\ 2001, \mnras, 323, 785

\bibitem[{{Caselli} {et~al.}(2019){Caselli}, {Pineda}, {Zhao}, {Walmsley},
  {Keto}, {Tafalla}, {Chac{\'o}n-Tanarro}, {Bourke}, {Friesen}, \&
  {Galli}}]{cas2019}
{Caselli}, P., {et~al.} 2019, \apj, 874, 89

\bibitem[{{Cesaroni} {et~al.}(2017){Cesaroni}, {S{\'a}nchez-Monge},
  {Beltr{\'a}n}, {Johnston}, {Maud}, {Moscadelli}, {Mottram}, {Ahmadi},
  {Allen}, {Beuther}, {Csengeri}, {Etoka}, {Fuller}, {Galli},
  {Galv{\'a}n-Madrid}, {Goddi}, {Henning}, {Hoare}, {Klaassen}, {Kuiper},
  {Kumar}, {Lumsden}, {Peters}, {Rivilla}, {Schilke}, {Testi}, {van der Tak},
  {Vig}, {Walmsley}, \& {Zinnecker}}]{ces2017}
{Cesaroni}, R., {et~al.} 2017, \aap, 602, A59

\bibitem[{{DeRose} {et~al.}(2009){DeRose}, {Bourke}, {Gutermuth}, {Wolk},
  {Megeath}, {Alves}, \& {N{\"u}rnberger}}]{der2009}
{DeRose}, K.~L., {Bourke}, T.~L., {Gutermuth}, R.~A., {Wolk}, S.~J., {Megeath},
  S.~T., {Alves}, J., \& {N{\"u}rnberger}, D. 2009, \aj, 138, 33

\bibitem[{{Dunham} and {Vorobyov}(2012){Dunham} \& {Vorobyov}}]{dun2012}
{Dunham}, M.~M., \& {Vorobyov}, E.~I. 2012, \apj, 747, 52

\bibitem[Dunham et al.(2016)]{dun2016} Dunham, M.~M., et al.\ 2016, \apj, 823, 160

\bibitem[{{Elmegreen} and {Lada}(1977){Elmegreen} \& {Lada}}]{elm1977}
{Elmegreen}, B.~G., \& {Lada}, C.~J. 1977, \apj, 214, 725

\bibitem[{{Frerking}, {Langer}, and {Wilson}(1982){Frerking}, {Langer}, \&
  {Wilson}}]{fre1982}
{Frerking}, M.~A., {Langer}, W.~D., \& {Wilson}, R.~W. 1982, \apj, 262, 590

\bibitem[{{Fukui} {et~al.}(2016){Fukui}, {Torii}, {Ohama}, {Hasegawa},
  {Hattori}, {Sano}, {Ohashi}, {Fujii}, {Kuwahara}, {Mizuno}, {Dawson},
  {Yamamoto}, {Tachihara}, {Okuda}, {Onishi}, \& {Mizuno}}]{fuk2016}
{Fukui}, Y., {et~al.} 2016, \apj, 820, 26

\bibitem[{{Fukui} {et~al.}(2018){Fukui}, {Torii}, {Hattori}, {Nishimura},
  {Ohama}, {Shimajiri}, {Shima}, {Habe}, {Sano}, {Kohno}, {Yamamoto},
  {Tachihara}, \& {Onishi}}]{fuk2018}
{Fukui}, Y., {et~al.} 2018, \apj, 859, 166

\bibitem[Fukui et al.(2019b)]{fuk2019_N159} Fukui, Y., et al.\ 2019b, \apj, 886, 14

\bibitem[Fukui et al.(2021a)]{fuk2021_cmf} Fukui, Y., Inoue, T., Hayakawa, T., et al.\ 2020, \pasj, doi:10.1093/pasj/psaa079

\bibitem[Fukui et al.(2021b)]{fuk2021b} Fukui, Y., Habe, A., Inoue, T., Enokiya, R. \& Tachihara, K.\ 2020, \pasj, submitted (arXiv:2009.05077)

\bibitem[{{Furniss}, {Jennings}, and {Moorwood}(1975){Furniss}, {Jennings}, \&
  {Moorwood}}]{fur1975}
{Furniss}, I., {Jennings}, R.~E., \& {Moorwood}, A.~F.~M. 1975, \apj, 202, 400

\bibitem[{{Guzm{\'a}n} {et~al.}(2014){Guzm{\'a}n}, {Garay}, {Rodr{\'{\i}}guez},
  {Moran}, {Brooks}, {Bronfman}, {Nyman}, {Sanhueza}, \& {Mardones}}]{guz2014}
{Guzm{\'a}n}, A.~E., {et~al.} 2014, \apj, 796, 117

\bibitem[{{Hirota} {et~al.}(2017){Hirota}, {Machida}, {Matsushita}, {Motogi},
  {Matsumoto}, {Kim}, {Burns}, \& {Honma}}]{hir2017}
{Hirota}, T., {Machida}, M.~N., {Matsushita}, Y., {Motogi}, K., {Matsumoto},
  N., {Kim}, M.~K., {Burns}, R.~A., \& {Honma}, M. 2017, Nature Astronomy, 1,
  0146

\bibitem[{{Hosokawa} {et~al.}(2016){Hosokawa}, {Hirano}, {Kuiper}, {Yorke},
  {Omukai}, \& {Yoshida}}]{hos2016}
{Hosokawa}, T., {Hirano}, S., {Kuiper}, R., {Yorke}, H.~W., {Omukai}, K., \&
  {Yoshida}, N. 2016, \apj, 824, 119

\bibitem[{{Hosokawa} and {Inutsuka}(2006){Hosokawa} \& {Inutsuka}}]{hos2006}
{Hosokawa}, T., \& {Inutsuka}, S.-i. 2006, \apj, 646, 240

\bibitem[{{Hosokawa} and {Omukai}(2009){Hosokawa} \& {Omukai}}]{hos2009}
{Hosokawa}, T., \& {Omukai}, K. 2009, \apj, 691, 823

\bibitem[{{Hosokawa}, {Yorke}, and {Omukai}(2010){Hosokawa}, {Yorke}, \&
  {Omukai}}]{hos2010}
{Hosokawa}, T., {Yorke}, H.~W., \& {Omukai}, K. 2010, \apj, 721, 478

\bibitem[{{Huchtmeier}(1974){Huchtmeier}}]{huc1974}
{Huchtmeier}, W. 1974, \aap, 32, 335

\bibitem[{{Inoue} and {Fukui}(2013){Inoue} \& {Fukui}}]{ino2013}
{Inoue}, T., \& {Fukui}, Y. 2013, \apjl, 774, L31

\bibitem[{{Inoue} {et~al.}(2018){Inoue}, {Hennebelle}, {Fukui}, {Matsumoto},
  {Iwasaki}, \& {Inutsuka}}]{ino2018}
{Inoue}, T., {Hennebelle}, P., {Fukui}, Y., {Matsumoto}, T., {Iwasaki}, K., \&
  {Inutsuka}, S.-i. 2018, \pasj, 70, S53

\bibitem[{{Izumi} {et~al.}(2019){Izumi}, {Fukui}, {Tokuda}, {Saigo}, {Harada},
  {Tachihara}, {Tsuge}, {Inoue}, {Torii}, {Nishimura}, \&
  {Zahorecz}}]{izu2019_pdf}
{Izumi}, N., {et~al.} 2019, in preparation

\bibitem[{{Johnston} {et~al.}(2015){Johnston}, {Robitaille}, {Beuther}, {Linz},
  {Boley}, {Kuiper}, {Keto}, {Hoare}, \& {van Boekel}}]{joh2015}
{Johnston}, K.~G., {et~al.} 2015, \apjl, 813, L19

\bibitem[{{Krumholz} {et~al.}(2010){Krumholz}, {Cunningham}, {Klein}, \&
  {McKee}}]{kur2010}
{Krumholz}, M.~R., {Cunningham}, A.~J., {Klein}, R.~I., \& {McKee}, C.~F. 2010,
  \apj, 713, 1120

\bibitem[{{Krumholz}, {Klein}, and {McKee}(2007){Krumholz}, {Klein}, \&
  {McKee}}]{kur2007}
{Krumholz}, M.~R., {Klein}, R.~I., \& {McKee}, C.~F. 2007, \apj, 656, 959

\bibitem[{{Krumholz} {et~al.}(2009){Krumholz}, {Klein}, {McKee}, {Offner}, \&
  {Cunningham}}]{kur2009}
{Krumholz}, M.~R., {Klein}, R.~I., {McKee}, C.~F., {Offner}, S.~S.~R., \&
  {Cunningham}, A.~J. 2009, Science, 323, 754

\bibitem[Krumholz et al.(2012)]{kur2012} Krumholz, M.~R., Klein, R.~I., \& McKee, C.~F.\ 2012, \apj, 754, 71

\bibitem[{{Kuhn}, {Getman}, and {Feigelson}(2015){Kuhn}, {Getman}, \&
  {Feigelson}}]{kuh2015}
{Kuhn}, M.~A., {Getman}, K.~V., \& {Feigelson}, E.~D. 2015, \apj, 802, 60

\bibitem[{{Kuiper} and {Hosokawa}(2018){Kuiper} \& {Hosokawa}}]{kui2018}
{Kuiper}, R., \& {Hosokawa}, T. 2018, \aap, 616, A101

\bibitem[{{Kuiper} {et~al.}(2010){Kuiper}, {Klahr}, {Beuther}, \&
  {Henning}}]{kui2010}
{Kuiper}, R., {Klahr}, H., {Beuther}, H., \& {Henning}, T. 2010, \apj, 722,
  1556

\bibitem[{{Kuiper} {et~al.}(2011){Kuiper}, {Klahr}, {Beuther}, \&
  {Henning}}]{kui2011}
{Kuiper}, R., {Klahr}, H., {Beuther}, H., \& {Henning}, T. 2011, \apj, 732, 20

\bibitem[{{Leung}, {Herbst}, and {Huebner}(1984){Leung}, {Herbst}, \&
  {Huebner}}]{leu1984}
{Leung}, C.~M., {Herbst}, E., \& {Huebner}, W.~F. 1984, \apjs, 56, 231

\bibitem[MacLaren et al.(1988)]{Mac1988} MacLaren, I., Richardson, K.~M., \& Wolfendale, A.~W.\ 1988, \apj, 333, 821

\bibitem[{{Matsushita} {et~al.}(2019){Matsushita}, {Takahashi}, {Machida}, \&
  {Tomisaka}}]{mat2019}
{Matsushita}, Y., {Takahashi}, S., {Machida}, M.~N., \& {Tomisaka}, K. 2019,
  \apj, 871, 221

\bibitem[{{Maud} {et~al.}(2015){Maud}, {Moore}, {Lumsden}, {Mottram},
  {Urquhart}, \& {Hoare}}]{mau2015a}
{Maud}, L.~T., {Moore}, T.~J.~T., {Lumsden}, S.~L., {Mottram}, J.~C.,
  {Urquhart}, J.~S., \& {Hoare}, M.~G. 2015, \mnras, 453, 645

\bibitem[{{McKee} and {Tan}(2003){McKee} \& {Tan}}]{mck2003}
{McKee}, C.~F., \& {Tan}, J.~C. 2003, \apj, 585, 850

\bibitem[{{McMullin} {et~al.}(2007){McMullin}, {Waters}, {Schiebel}, {Young},
  \& {Golap}}]{mcm2007}
{McMullin}, J.~P., {Waters}, B., {Schiebel}, D., {Young}, W., \& {Golap}, K.
  2007, in Astronomical Society of the Pacific Conference Series, Vol. 376,
  Astronomical Data Analysis Software and Systems XVI, ed. R.~A. {Shaw},
  F.~{Hill}, \& D.~J. {Bell}, 127

\bibitem[{{Meyer} {et~al.}(2018){Meyer}, {Kuiper}, {Kley}, {Johnston}, \&
  {Vorobyov}}]{mey2018}
{Meyer}, D.~M.~A., {Kuiper}, R., {Kley}, W., {Johnston}, K.~G., \& {Vorobyov},
  E. 2018, \mnras, 473, 3615

\bibitem[{{Meyer} {et~al.}(2017){Meyer}, {Vorobyov}, {Kuiper}, \&
  {Kley}}]{mey2017}
{Meyer}, D.~M.~A., {Vorobyov}, E.~I., {Kuiper}, R., \& {Kley}, W. 2017, \mnras,
  464, L90

\bibitem[{{Motogi} {et~al.}(2019){Motogi}, {Hirota}, {Machida}, {Yonekura},
  {Honma}, {Takakuwa}, \& {Matsushita}}]{mot2019}
{Motogi}, K., {Hirota}, T., {Machida}, M.~N., {Yonekura}, Y., {Honma}, M.,
  {Takakuwa}, S., \& {Matsushita}, S. 2019, \apj, 877, L25

\bibitem[{{Motogi} {et~al.}(2017){Motogi}, {Hirota}, {Sorai}, {Yonekura},
  {Sugiyama}, {Honma}, {Niinuma}, {Hachisuka}, {Fujisawa}, \&
  {Walsh}}]{mot2017}
{Motogi}, K., {et~al.} 2017, \apj, 849, 23

\bibitem[{{Motte} {et~al.}(2018){Motte}, {Nony}, {Louvet}, {Marsh}, {Bontemps},
  {Whitworth}, {Men'shchikov}, {Nguyen Luong}, {Csengeri}, {Maury}, {Gusdorf},
  {Chapillon}, {K{\"o}nyves}, {Schilke}, {Duarte-Cabral}, {Didelon}, \&
  {Gaudel}}]{mot2018}
{Motte}, F., {et~al.} 2018, Nature Astronomy, 2, 478

\bibitem[{{Myers}(2009){Myers}}]{mye2009}
{Myers}, P.~C. 2009, \apj, 700, 1609

\bibitem[{{Nakano}(1989){Nakano}}]{nak1989}
{Nakano}, T. 1989, \apj, 345, 464

\bibitem[{{Offner} {et~al.}(2011){Offner}, {Lee}, {Goodman}, \&
  {Arce}}]{off2011}
{Offner}, S. S.~R., {Lee}, E.~J., {Goodman}, A.~A., \& {Arce}, H. 2011, \apj,
  743, 91

\bibitem[{{Ohashi} {et~al.}(2014){Ohashi}, {Saigo}, {Aso}, {Aikawa},
  {Koyamatsu}, {Machida}, {Saito}, {Takahashi}, {Takakuwa}, {Tomida},
  {Tomisaka}, \& {Yen}}]{oha2014}
{Ohashi}, N., {et~al.} 2014, \apj, 796, 131

\bibitem[Ohashi et al.(2016)]{oha2016} Ohashi, S., Sanhueza, P., Chen, H.-R.~V., et al.\ 2016, \apj, 833, 209

\bibitem[{{Ohashi} {et~al.}(2018){Ohashi}, {Sanhueza}, {Sakai}, {Kand ori},
  {Choi}, {Hirota}, {Nguyen-Luong}, \& {Tatematsu}}]{oha2018}
{Ohashi}, S., {Sanhueza}, P., {Sakai}, N., {Kand ori}, R., {Choi}, M.,
  {Hirota}, T., {Nguyen-Luong}, Q., \& {Tatematsu}, K. 2018, \apj, 856, 147

\bibitem[Onishi et al.(2002)]{oni2002} Onishi, T., Mizuno, A., Kawamura, A., et al.\ 2002, \apj, 575, 950

\bibitem[{{Ossenkopf} and {Henning}(1994){Ossenkopf} \& {Henning}}]{oss1994}
{Ossenkopf}, V., \& {Henning}, T. 1994, \aap, 291, 943

\bibitem[{{Peters} {et~al.}(2010a){Peters}, {Banerjee}, {Klessen}, {Mac Low},
  {Galv{\'a}n-Madrid}, \& {Keto}}]{pet2010a}
{Peters}, T., {Banerjee}, R., {Klessen}, R.~S., {Mac Low}, M.-M.,
  {Galv{\'a}n-Madrid}, R., \& {Keto}, E.~R. 2010, \apj, 711, 1017

\bibitem[Peters et al.(2010b)]{pet2010b} Peters, T., Klessen, R.~S., Mac Low, M.-M., et al.\ 2010, \apj, 725, 134

\bibitem[{{S{\'a}nchez-Monge} {et~al.}(2013){S{\'a}nchez-Monge}, {Cesaroni},
  {Beltr{\'a}n}, {Kumar}, {Stanke}, {Zinnecker}, {Etoka}, {Galli}, {Hummel},
  {Moscadelli}, {Preibisch}, {Ratzka}, {van der Tak}, {Vig}, {Walmsley}, \&
  {Wang}}]{san2013}
{S{\'a}nchez-Monge}, {\'A}., {et~al.} 2013, \aap, 552, L10

\bibitem[Shima et al.(2017)]{shima2017} Shima, K., Tasker, E.~J., \& Habe, A.\ 2017, \mnras, 467, 512

\bibitem[{{Stahler}, {Palla}, and {Salpeter}(1986){Stahler}, {Palla}, \&
  {Salpeter}}]{sta1986}
{Stahler}, S.~W., {Palla}, F., \& {Salpeter}, E.~E. 1986, \apj, 302, 590

\bibitem[Spaans \& van Dishoeck(1997)]{spa1997} Spaans, M., \& van Dishoeck, E.~F.\ 1997, \aap, 323, 953

\bibitem[Tachihara et al.(2018)]{tac2018} Tachihara, K., Fukui, Y., Hayakawa, T., et al.\ 2018, arXiv e-prints, arXiv:1811.02224

\bibitem[Tachihara et al.(2002)]{tac2002} Tachihara, K., Onishi, T., Mizuno, A., et al.\ 2002, \aap, 385, 909

\bibitem[{{Tan} {et~al.}(2014){Tan}, {Beltr{\'a}n}, {Caselli}, {Fontani},
  {Fuente}, {Krumholz}, {McKee}, \& {Stolte}}]{tan2014}
{Tan}, J.~C., {Beltr{\'a}n}, M.~T., {Caselli}, P., {Fontani}, F., {Fuente}, A.,
  {Krumholz}, M.~R., {McKee}, C.~F., \& {Stolte}, A. 2014, Protostars and
  Planets VI, 149

\bibitem[{{Tokuda} {et~al.}(2014){Tokuda}, {Onishi}, {Saigo}, {Kawamura},
  {Fukui}, {Matsumoto}, {Inutsuka}, {Machida}, {Tomida}, \&
  {Tachihara}}]{tok2014}
{Tokuda}, K., {et~al.} 2014, \apj, 789, L4

\bibitem[{{Tokuda} {et~al.}(2016){Tokuda}, {Onishi}, {Matsumoto}, {Saigo},
  {Kawamura}, {Fukui}, {Inutsuka}, {Machida}, {Tomida}, \&
  {Tachihara}}]{tok2016}
{Tokuda}, K., {et~al.} 2016, \apj, 826, 26

\bibitem[{{Tokuda} {et~al.}(2019a){Tokuda}, {Fukui}, {Harada}, {Saigo},
  {Tachihara}, {Tsuge}, {Inoue}, {Torii}, {Nishimura}, \& {Zahorecz}}]{tok2019a}
{Tokuda}, K., {et~al.} 2019a, \apj, 886, 15

\bibitem[{{Tokuda} {et~al.}(2019b){Tokuda}, {Tachihara}, {Saigo}, {Andr{\'e}},
  {Miyamoto}, {Zahorecz}, {Inutsuka}, {Matsumoto}, {Takashima}, \&
  {Machida}}]{tok2019b}
{Tokuda}, K., {et~al.} 2019b, \pasj, 71, 73

\bibitem[Tokuda et al.(2020)]{tok2020} Tokuda, K., et al.\ 2020, \apj, 899, 10

\bibitem[Torii et al.(2011)]{tor2011} Torii, K., et al.\ 2011, \apj, 738, 46

\bibitem[{{Torii} {et~al.}(2015){Torii}, {Hasegawa}, {Hattori}, {Sano},
  {Ohama}, {Yamamoto}, {Tachihara}, {Soga}, {Shimizu}, {Okuda}, {Mizuno},
  {Onishi}, {Mizuno}, \& {Fukui}}]{tor2015}
{Torii}, K., {et~al.} 2015, \apj, 806, 7

\bibitem[{{Torii} {et~al.}(2017{\natexlab{a}}){Torii}, {Hattori}, {Hasegawa},
  {Ohama}, {Yamamoto}, {Tachihara}, {Tokuda}, {Onishi}, {Hattori}, \&
  {Ishihara}}]{tor2017a}
{Torii}, K., {et~al.} 2017{\natexlab{a}}, \apj, 840, 111

\bibitem[{{Torii} {et~al.}(2017{\natexlab{b}}){Torii}, {Hattori}, {Hasegawa},
  {Ohama}, {Haworth}, {Shima}, {Habe}, {Tachihara}, {Mizuno}, {Onishi},
  {Mizuno}, \& {Fukui}}]{tor2017}
{Torii}, K., {et~al.} 2017{\natexlab{b}}, \apj, 835, 142

\bibitem[{{Vigil}(2004){Vigil}}]{vig2004}
{Vigil}, M. 2004, Master's thesis, Massachusetts Institute of Technology

\bibitem[Wang et al.(2010)]{wan2010} Wang, P., Li, Z.-Y., Abel, T., et al.\ 2010, \apj, 709, 27

\bibitem[{{Williams}, {de Geus}, and {Blitz}(1994){Williams}, {de Geus}, \&
  {Blitz}}]{clumpfind_wil1994}
{Williams}, J.~P., {de Geus}, E.~J., \& {Blitz}, L. 1994, \apj, 428, 693

\bibitem[{{Wilson} and {Rood}(1994){Wilson} \& {Rood}}]{wil1994}
{Wilson}, T.~L., \& {Rood}, R. 1994, \araa, 32, 191

\bibitem[{{Winston} {et~al.}(2011){Winston}, {Wolk}, {Bourke}, {Megeath},
  {Gutermuth}, \& {Spitzbart}}]{win2011}
{Winston}, E., {Wolk}, S.~J., {Bourke}, T.~L., {Megeath}, S.~T., {Gutermuth},
  R., \& {Spitzbart}, B. 2011, \apj, 743, 166

\bibitem[{{Wolfire} and {Cassinelli}(1986){Wolfire} \& {Cassinelli}}]{wol1986}
{Wolfire}, M.~G., \& {Cassinelli}, J.~P. 1986, \apj, 310, 207

\bibitem[{{Wolk}, {Bourke}, and {Vigil}(2008){Wolk}, {Bourke}, \&
  {Vigil}}]{wol2008}
{Wolk}, S.~J., {Bourke}, T.~L., \& {Vigil}, M. 2008, {The Embedded Massive Star
  Forming Region RCW 38}, ed. B.~{Reipurth}, 124

\bibitem[{{Wolk} {et~al.}(2006){Wolk}, {Spitzbart}, {Bourke}, \&
  {Alves}}]{wol2006}
{Wolk}, S.~J., {Spitzbart}, B.~D., {Bourke}, T.~L., \& {Alves}, J. 2006, \aj,
  132, 1100

\bibitem[{{Zhang} {et~al.}(2019){Zhang}, {Tan}, {Sakai}, {Tanaka}, {De Buizer},
  {Liu}, {Beltr{\'a}n}, {Kratter}, {Mardones}, \& {Garay}}]{zha2019}
{Zhang}, Y., {et~al.} 2019, \apj, 873, 73

\end{thebibliography}

%\appendix

\clearpage
\begin{table}
  \tbl{Spectral parameters of the ALMA observations}{
 \small
  \begin{tabular}{ccccc}
  \hline
  \hline
     Line/Continuum  & Band & Frequency & Spectral resolution & Array \\
     && (GHz) & (kHz) \\
%     (1) & (2) & (3) & (4) & (5) & (6) &  & (7) & (8) & (9) && (10) & (11) && (12)\\
    \hline
    Continuum &6 & 233.000 & 31.250 (MHz) & 12-m, ACA \\
    Continuum &7 & 341.000 & 31.250 (MHz) & ACA \\
    \hline
    $^{12}$CO\,$J$=2--1 &6 &  230.538 & 141 & 12-m, ACA \\
    $^{13}$CO\,$J$=2--1 &6 &  220.399 & 141 & 12-m, ACA \\
    C$^{18}$O\,$J$=2--1 &6 &  219.560 & 141 & 12-m, ACA \\
    SiO\,$v$=0\,$J$=5--4 & 6 & 217.105 & 244 & 12-m, ACA\\
    H30$\alpha$ & 6 & 231.901 & 141 & 12-m, ACA \\
    \hline
    $^{13}$CO\,$J$=3--2 &7 &  330.588 & 244 & ACA \\
    C$^{18}$O\,$J$=3--2 &7 &  329.331 & 244 & ACA \\
    CS\,$J$=7--6 & 7 & 342.883 & 244 & ACA\\
  \hline
  \end{tabular} }\label{tab:1}
   \begin{tabnote}
   %memo
   \end{tabnote}
\end{table}

\begin{table}
  \tbl{List of the ALMA datasets used in this paper}{
 \small
  \begin{tabular}{ccccccccccccccc}
  \hline
  \hline
   Image &  Array & Spectral resolution & Beam & Noise$^{a}$ & Name$^{b}$\\
     && (km\,s$^{-1}$) & ($''$) & (mJy\,beam$^{-1}$)  \\
    \hline
   Continuum (233-GHz) 				& 12-m	 		& ... 		& $0.17\times0.15$ 		& 0.45	& 233-GHz \\
   $^{12}$CO\,$J$=2--1 				& 12-m	 		& 1.0 	& $0.50\times0.49$		& 4.3		&  $^{12}$CO \\
   C$^{18}$O\,$J$=2--1 				& 12-m	 		& 0.5 	& $0.23\times0.21$		& 19.6	& C$^{18}$O \\
   H30$\alpha$     & 12-m          & 2.0   & $0.23\times0.20$      & 8.2 &   H30$\alpha$ \\
   %SiO\,$v$=0\,$J$=2--1 				& 12m only 		& 1.0 	& $0.23\times0.21$		& 12.3	& SiO \\
   \hline
   $^{13}$CO\,$J$=2--1				& 12-m + ACA 	 	& 0.2 	& $1.90\times1.55$		& 123.8	& $^{13}$CO$_{\rm all}$  \\
   C$^{18}$O\,$J$=2--1				& 12-m + ACA 	 	& 0.2 	& $2.13\times1.74$		& 93.1	& C$^{18}$O$_{\rm all}$  \\   
   \hline
   Continuum (233-GHz)				& ACA			& ...		& $7.25\times5.29$		& 15.3	& 233-GHz$_{\rm ACA}$ \\
   Continuum (341-GHz)				& ACA			& ...		& $4.75\times3.65$		& 37.2	& 341-GHz$_{\rm ACA}$ \\
   $^{12}$CO\,$J$=2--1				& ACA			& 0.25	& $7.3\times5.2$		& 167	& $^{12}$CO$_{\rm ACA}$\\
   C$^{18}$O\,$J$=3--2				& ACA 			& 0.25	& $5.0\times3.6$		& 215	& C$^{18}$O$_{\rm ACA}$\\
   H30$\alpha$						& ACA 			& 0.25	& $7.2\times5.3$		& 150	& H30$\alpha_{\rm ACA}$\\
   %CS\,$J$=7--6						& ACA + TP		& 0.25	& $4.8\times3.4$		& 218	& CS$_{\rm ACA}$\\   
   \hline
  \end{tabular} }\label{tab:2}
   \begin{tabnote}
   $^{(a)}$ The noise is given per channel for line data cubes. $^{(b)}$ The shortened name of the dataset used in this paper. %$^{(c)}$ The 12m array data was taperred.
   \end{tabnote}
\end{table}

\begin{table}
  \tbl{Physical properties of the C$^{18}$O condensations}{
 \small
  \begin{tabular}{ccccccccccccccc}
  \hline
  \hline
\# & R.A. (J2000)  & decl. (J2000) & $r$  & {\bf $\sigma_v$} & peak $N_{\rm H_2}$ & $M_{\rm H_2}$ & $n_{\rm H_2}$ & $M_{\rm vir}$ \\ 
 & & & (pc)/(AU) & (km\,s$^{-1}$) & ($\times 10^{23}$\,cm$^{-2}$) & ($M_\odot$) & ($\times 10^{7}$\,cm$^{-3}$) & ($M_\odot$) \\ 
\hline
1  & 8:59:06.60 & -47:29:27.7 & 0.021/4370 & 1.1 & 5.5--5.6 & 28--34 & 1.1--1.3 & 28 \\ 
2  & 8:59:06.13 & -47:29:37.7 & 0.019/4010 & 1.2 & 2.4--2.6 & 20--24 & 1.0--1.2 & 33 \\ 
3  & 8:59:04.70 & -47:29:58.3 & 0.015/3090 & 0.8 & 3.3--4.1 & 13--15 & 1.4--1.6 & 11 \\ 
4  & 8:59:08.35 & -47:30:12.0 & 0.006/1340 & 0.8 & 2.6--3.2 & 2--2 & 3.0--3.3 & 4 \\ 
5  & 8:59:01.42 & -47:30:12.0 & 0.014/2980 & 1.0 & 1.1--1.3 & 4--6 & 0.6--0.7 & 16 \\ 
6  & 8:59:02.33 & -47:30:19.5 & 0.019/3990 & 1.6 & 2.6--3.1 & 23--27 & 1.1--1.3 & 60 \\ 
7  & 8:59:03.19 & -47:30:23.0 & 0.020/4210 & 1.1 & 3.5--3.5 & 27--31 & 1.1--1.3 & 30 \\ 
8  & 8:59:01.34 & -47:30:29.7 & 0.017/3440 & 0.9 & 2.8--3.4 & 17--20 & 1.3--1.5 & 17 \\ 
9  & 8:59:01.54 & -47:30:38.0 & 0.024/4880 & 1.0 & 4.2--5.2 & 46--52 & 1.2--1.4 & 30 \\ 
10 & 8:59:02.58 & -47:30:41.5 & 0.016/3280 & 1.3 & 3.8--4.2 & 17--19 & 1.5--1.7 & 30 \\ 
11 & 8:58:59.86 & -47:30:41.7 & 0.019/4000 & 1.3 & 2.5--3.2 & 19--22 & 0.9--1.1 & 36 \\ 
12 & 8:59:02.90 & -47:30:44.2 & 0.014/2780 & 1.0 & 4.4--6.7 & 18--18 & 2.6--2.6 & 14 \\ 
13 & 8:59:03.69 & -47:30:44.8 & 0.015/3160 & 1.6 & 10.2--10.4 & 27--34 & 2.6--3.4 & 44 \\ 
14 & 8:59:02.48 & -47:30:47.7 & 0.012/2410 & 0.7 & 2.8--3.6 & 8--8 & 1.8--2.0 & 6 \\ 
15 & 8:59:00.55 & -47:30:53.2 & 0.018/3790 & 1.2 & 3.7--4.6 & 23--27 & 1.3--1.5 & 30 \\ 
16 & 8:58:59.96 & -47:30:57.0 & 0.013/2750 & 1.4 & 4.4--4.4 & 14--16 & 2.2--2.5 & 29 \\ 
17 & 8:59:01.29 & -47:30:57.2 & 0.012/2550 & 0.8 & 2.5--3.1 & 9--10 & 1.8--1.9 & 8 \\ 
18 & 8:59:00.08 & -47:31:00.0 & 0.011/2250 & 1.3 & 5.2--8.2 & 10--15 & 2.7--4.1 & 21 \\ 
19 & 8:59:00.63 & -47:31:00.7 & 0.011/2260 & 0.9 & 4.0--4.5 & 9--11 & 2.6--3.0 & 11 \\ 
20 & 8:59:01.12 & -47:31:02.5 & 0.010/2120 & 0.8 & 3.4--5.9 & 7--10 & 2.5--3.5 & 7 \\ 
21 & 8:59:00.41 & -47:31:06.0 & 0.010/2070 & 1.4 & 3.3--7.7 & 6--11 & 2.1--4.1 & 22 \\ 
   \hline
  \end{tabular} }\label{tab:1.5}
   \begin{tabnote}
   $^{(a)}$ $\sigma_v$ is the standard deviation of the velocity.
   $^{(b)}$ $n_{\rm H_2}$ was calculated from $M_{\rm H_2}$ assuming a sphere with a radius of $r$.
   \end{tabnote}
\end{table}

\begin{table}
  \tbl{Measured parameters of Source\,A and Source\,B}{
 \scriptsize
  \begin{tabular}{cccccccccccc}
  \hline
  \hline
     \multirow{2}{*}{ Source } & \multirow{2}{*}{ R.A. (J2000) } & \multirow{2}{*}{ decl. (J2000) } & \multicolumn{2}{c}{341-GHz$_{\rm ACA}$ ($4''.75\times3''.65$)} &&  \multicolumn{2}{c}{233-GHz ($0''.17\times0''.15$)} \\
     \cline{4-5} \cline{7-8}
      && & $S_{\rm peak}$ & $S_{\rm int}$$^{a}$ && $S_{\rm peak}$&$S_{\rm int}$$^{a}$ \\
     & ($^{\rm h\,m\,s}$) & ($^\circ$\,$'$\,$''$) & (Jy\,beam$^{-1}$) & (Jy) && (mJy\,beam$^{-1}$) & (Jy) & \\
     \hline
     A & 8:59:3.740 & -47:30:44.832 &   3.1   &  4.9(0.6)   && 24.9 & 0.24(0.07) \\
     B &8:59:3.013  & -47:30:21.982 &  1.4    &  2.5(0.8)   && 48.8 & 0.11(0.05)  \\
     \hline
  \end{tabular} }\label{tab:2.5}
   \begin{tabnote}
   $^{(a)}$ The aperture radii of the photometry for measuring $S_{\rm int}$ were set to $6''.5$ and $0''.5$ for the 341-GHz$_{\rm ACA}$ image and the 233-GHz image, respectively. In the 341-GHz$_{\rm ACA}$ image, the local background level of each source was estimated around the source using a circular annulus with inner and outer radii of $6''.5$ and $1.5\times6''.5 = 9''.8$, respectively. The median value of the pixels within the annulus was applied as the local background level, with the standard deviation applied as the uncertainty of the photometry. The derived uncertainty is shown in parentheses.
   \end{tabnote}
\end{table}

\begin{table}
  \tbl{Physical properties of Source\,A and Source\,B}{
 \scriptsize
  \begin{tabular}{cccccccccccc}
  \hline
  \hline
     \multirow{2}{*}{ Source }  & \multicolumn{3}{c}{341-GHz$_{\rm ACA}$ ($4''.75\times3''.65$)} &&  \multicolumn{3}{c}{233-GHz ($0''.17\times0''.15$)} \\
     \cline{2-4} \cline{6-8}
     &  $r_{\rm dust}$$^{a}$ & peak $N_{\rm H_2}$$^{b}$ & $M_{\rm H_2}$$^{b}$ &&  $r_{\rm dust}$$^{a}$ & peak $N_{\rm H_2}$$^{b}$ & $M_{\rm H_2}$$^{b}$\\
     &  (pc) & ($\times10^{23}$\,cm$^{-2}$) & ($M_\odot$) && (AU) & ($\times10^{24}$\,cm$^{-2}$) & ($M_\odot$)\\
     \hline
     A &    0.02 & 10.7 -- 26.8 & 32(3.9) -- 81(9.9) &&    460 & 6 & 5(3.8) \\
     B &    0.03 & 4.9 -- 12.2  & 16(5.1) -- 41(13.1)  &&  200  & 6 & 2(1.2) \\
     \hline
  \end{tabular} }\label{tab:3}
   \begin{tabnote}
   $^{(a)}$ $r_{\rm dust}$ was measured at the half maximum.
   $^{(b)}$ The $N_{\rm H_2}$ and $M_{\rm H_2}$ were estimated at the optically-thin limit. The $T_{\rm d}$ of 20 and 40\,K were assumed in the $N_{\rm H_2}$ and $M_{\rm H_2}$ estimated from the 341-GHz$_{\rm ACA}$ image to constrain $M_{\rm H_2}$ ranges of the sources, whereas in the 233-GHz image, 100\,K (see the text) was used as $T_{\rm d}$ for Source\,A and Source\,B, respectively (Figure\,\ref{233}). The statistical uncertainty of the derived $M_{\rm H_2}$ is shown in parentheses. Note that systematic errors arising from uncertainties of the distance measurement and the assumption of gas-to-dust ratio/opacity give factor of $\sim$2 uncertainty further.
   \end{tabnote}
\end{table}

\begin{table}
  \tbl{Physical properties of the outflow lobes}{
 \small
  \begin{tabular}{lcccccccccccccc}
  \hline
  \hline
   Lobes & $v_{\rm max}$/sin$i$ & $l_{\rm max}$/cos$i$ & $t_{\rm dyn}$/tan$i$ & $M_{\rm lobe}$ & $\dot{M_{\rm lobe}}$//tan$i$ \\
   & (km\,s$^{-1}$) & (AU) & (yr) & ($M_\odot$) & ($M_\odot$\,yr$^{-1}$)\\ 
    \hline
    (Source\,A: $v_{\rm sys} = 2$\,km\,s$^{-1}$) \\
    \ \ \ Blueshifted lobe & 24 & 3200 & 630& 0.06 & $0.10\times10^{-3}$\\
    \ \ \ Redshifted lobe  & 38 & 9100 & 1100 & 0.4  & $0.36\times10^{-3}$ \\
    (Source\,B: $v_{\rm sys} = 5$\,km\,s$^{-1}$) \\
    \ \ \ Blueshifted lobe & 67 & 10600 & 750 & 1.3  & $1.7\times10^{-3}$\\
    \ \ \ Redshifted lobe  & 45 & ... & ...& 0.3  & ... \\    
   \hline
  \end{tabular} }\label{tab:4}
   \begin{tabnote}
	$v_{\rm max}$ is the maximum velocity of the outflow lobe measured from the systemic velocity $v_{\rm sys}$ of the source, whereas $l_{\rm max}$ is the physical length of the lobes measured from the 233-GHz peak of the source. The dynamical timescale of the outflow $t_{\rm dyn}$ was calculated as $t_{\rm dyn} = l_{\rm max}/v_{\rm max}$. The H$_2$ mass of the outflow lobe was calculated using a CO-to-H$_2$ conversion factor of $2\times10^{20}$\,(K\,km\,s$^{-1}$)$^{-1}$\,cm$^{-2}$ \citep{bol2013}. The mass flow rate $\dot{M_{\rm lobe}}$ was measured as $M_{\rm flow}/t_{\rm dyn}$. The $l_{\rm max}$ of the redshifted lobe in Source\,B could not be measured.
   \end{tabnote}
\end{table}

\clearpage

\end{document}